\documentstyle[psfig]{mn}

\input{epsf}

\def\lsim{\mathrel{\hbox{\rlap{\hbox{\lower4pt\hbox{$\sim$}}}\hbox{$<$}}}}
\def\gsim{\mathrel{\hbox{\rlap{\hbox{\lower4pt\hbox{$\sim$}}}\hbox{$>$}}}}

\def\and {\rm {et al.} \rm} % Roman Font is the MNRAS/ApJ style these days

\begin{document}

\title[The 2dFGRS: The clustering of galaxy groups] 
{The 2dF Galaxy Redshift Survey: The clustering of galaxy groups}
\author[Padilla et al.]{
\parbox[t]{\textwidth}{
Nelson D.\ Padilla$^1$,
Carlton M.\ Baugh$^1$,
Vincent R.\ Eke$^1$,
Peder Norberg$^2$,
Shaun Cole$^1$, 
Carlos S.\ Frenk$^1$, 
Darren J.\ Croton$^3$,
Ivan K.\ Baldry$^4$,
Joss Bland-Hawthorn$^5$,
Terry Bridges$^{5,6}$, 
Russell Cannon$^5$, 
Matthew Colless$^{5,7}$, 
Chris Collins$^8$, 
Warrick Couch$^9$, 
Gavin Dalton$^{10,11}$,
Roberto De Propris$^7$,
Simon P.\ Driver$^7$, 
George Efstathiou$^{12}$, 
Richard S.\ Ellis$^{13}$, 
Karl Glazebrook$^4$, 
Carole Jackson$^7$,
Ofer Lahav$^{14}$, 
Ian Lewis$^{10}$, 
Stuart Lumsden$^{15}$, 
Steve Maddox$^{16}$,
Darren Madgwick$^{17}$,
John A.\ Peacock$^{18}$,
Bruce A.\ Peterson$^7$, 
Will Sutherland$^{18,12}$,
Keith Taylor$^{13}$}
\vspace*{6pt} \\ 
$^1$Department of Physics, University of Durham, South Road, 
    Durham DH1 3LE, UK.\\ 
$^{2}$ETHZ Instit\"ut f\"ur Astronomie, HPF G3.1, ETH Honggerberg, CH-8093
       Z\"urich, Switzerland.\\
$^{3}$Max-Planck-Instit\"ut f\"ur Astrophysik, D-85740 Garching, Germany.\\
$^4$Department of Physics \& Astronomy, Johns Hopkins University,
       Baltimore, MD 21118-2686, USA. \\
$^5$Anglo-Australian Observatory, P.O.\ Box 296, Epping, NSW 2111,
    Australia.\\  
$^6$Department of Physics, Queen's University, Kingston, Ontario, 
    K7L 3N6, Canada.\\
$^7$Research School of Astronomy \& Astrophysics, The Australian 
    National University, Weston Creek, ACT 2611, Australia. \\
$^8$Astrophysics Research Institute, Liverpool John Moores University,  
    Twelve Quays House, Birkenhead, L14 1LD, UK. \\
$^9$Department of Astrophysics, University of New South Wales, Sydney, 
    NSW 2052, Australia. \\
$^{10}$Department of Physics, University of Oxford, Keble Road, 
    Oxford OX1 3RH, UK. \\
$^{11}$Space Science and Technology Department, CCLRC Rutherford 
    Appleton Laboratory, Chilton, Didcot, Oxfordshire, OX11 0QX, UK.\\
$^{12}$Institute of Astronomy, University of Cambridge, Madingley Road,
    Cambridge CB3 0HA, UK. \\
$^{13}$Department of Astronomy, California Institute of Technology, 
    Pasadena, CA 91025, USA. \\
$^{14}$Department of Physics and Astronomy, University College London, 
Gower Street, London, WC1E 6BT, UK. \\
$^{15}$Department of Physics, University of Leeds, Woodhouse Lane,
       Leeds, LS2 9JT, UK. \\
$^{16}$School of Physics \& Astronomy, University of Nottingham,
       Nottingham NG7 2RD, UK. \\
$^{17}$Department of Astronomy, University of California, Berkeley, CA
       92720, USA. \\
$^{18}$Institute for Astronomy, University of Edinburgh, Royal Observatory, 
       Blackford Hill, Edinburgh EH9 3HJ, UK. \\
}

\maketitle

\begin{abstract}
We measure the clustering of galaxy groups in the 2dFGRS Percolation-Inferred 
Galaxy Group (2PIGG) catalogue. The 2PIGG sample has 28\,877 groups with 
at least two members. The clustering amplitude of the full 2PIGG catalogue is 
weaker than that of 2dFGRS galaxies, in agreement with theoretical 
predictions. 
We have subdivided the 2PIGG catalogue into samples that span a factor of 
$\approx 25$ in median total luminosity. 
Our correlation function measurements span an unprecedented range of 
clustering strengths, connecting the regimes probed by groups
fainter than $L_*$ galaxies and rich clusters.
There is a steady increase in clustering strength with group luminosity; 
the most luminous groups are ten times more strongly clustered 
than the full 2PIGG catalogue. 
We demonstrate that the 2PIGG results are in very good 
agreement with the clustering of groups expected in 
the $\Lambda$CDM model.
\end{abstract}


\section{Introduction}
\label{sec:intro}

Galaxy groups are important tracers of the matter distribution 
in the Universe, offering a powerful alternative to individual
galaxies. 
Part of the appeal of galaxy groups lies in the simplicity of their 
relation to the dark matter. 
Theoretically, each dark matter halo yields a single group of galaxies. 
This is a clear advantage over the use of galaxies to trace mass, 
as the occupation of haloes by galaxies depends upon halo mass 
(e.g. Benson et al. 2000; Berlind et al. 2003). 
Typical groups are less elitist than the richest clusters which only 
account for a few percent of the mass of the universe 
(e.g. Jenkins et al. 2001).
The measurement of the clustering amplitude of groups as a function 
of their mass is an important probe of hierarchical clustering 
(Governato et al. 1999; Colberg et al. 2000; Padilla \& Baugh 2002).

Previous attempts to measure the clustering of groups have been 
hampered by the small size of the samples. For example, 
Jing \& Zang (1988) measured the redshift space correlation 
function of a sample of 135 groups extracted from the CfA survey 
by Geller \& Huchra (1983). By the end of the 1990s, this situation had 
only improved slightly with an analysis of $\sim 500$ groups in the Updated 
Zwicky Catalogue by Merch\'{a}n, Maia \& Lambas (2000). 
Perhaps unsurprisingly, these early studies were inconclusive. 
One dispute concerned the relative strength of the auto-correlation 
function of groups and of galaxies. 
Kashlinsky (1987) produced an analytic argument suggesting that a 
sample of groups corresponding to haloes less massive than a 
characteristic mass $M_*$ (see Section 2.3 for a definition of $M_*$) should 
display a weaker clustering signal than that of their constituent galaxies. 
Similar conclusions can be reached from the calculation of the bias of 
dark matter haloes first carried out by Cole \& Kaiser (1989),
and subsequently by Mo \& White (1996) and Sheth, Mo \& Tormen (2001).
However, the first results comparing the clustering of groups and galaxies 
both confirmed (Jing \& Zang 1988; Maia \& da Costa 1990) and contradicted 
(Ramella, Geller \& Huchra 1990; Girardi, Boschin \& da Costa 2000) 
this theoretical prediction. 

An accurate measurement of the clustering of groups is now 
particularly timely as theoretical models have advanced 
to the point where detailed predictions can be 
made for the clustering of galactic systems. 
Techniques have been developed recently that allow high resolution 
N-body simulations of hierarchical clustering to be populated 
with galaxies using semi-analytic models (Kauffmann et al. 1999; 
Benson et al. 2000; Helly et al. 2003). 
The models make firm predictions for how the luminous baryonic 
component of the universe is partitioned between dark matter haloes.

A significant advance in statistical studies of group properties was 
made possible by the two-degree Field Galaxy Redshift Survey (2dFGRS; 
Colless et al. 2001; 2003). Merch\'{a}n \& Zandivarez (2002) used the 
2dFGRS 100k data release to construct a group catalogue, the 2dFGGC, 
consisting of 2198 groups with at least four members. The correlation 
function of the 2dFGGC in redshift space 
can be described by a power law with correlation 
length $s_0 = 8.9 \pm 0.3 h^{-1}$Mpc and slope $\gamma = -1.6 \pm 0.1$ 
(Zandivarez, Merch\'{a}n \& Padilla 2003). The clustering amplitude of 
2dFGGC groups is significantly higher than that measured for 2dFGRS galaxies, 
for which $s_0 = 6.82\pm 0.28h^{-1}$Mpc (Hawkins et al. 2003).  

The completion of the 2dFGRS has made it possible to construct a much 
larger group catalogue, the 2dFGRS Percolation Inferred Galaxy Group 
catalogue (hereafter the 2PIGG catalogue; Eke et al. 2004a). 
The gain is bigger than the simple factor of just over $2$ increase in 
the number of galaxy redshifts; the improved homogeneity 
and higher spectroscopic completeness means that group 
finding is far more efficient in the final 2dFGRS catalogue. To quantify 
this improvement, the number of groups with a minimum of four members 
in the 2PIGG catalogue is 7200, over three times the number of comparable 
groups detected by Merch\'{a}n \& Zandivarez (2002). 
Moreover, Eke et al. demonstrate that the group finding algorithm can 
be pushed further still to extract groups containing two or more members, 
which greatly increases the number of groups. 

In this paper, we study the clustering of groups in the 2PIGG catalogue.
The properties of the catalogue are summarized in Section 2, along with an 
explanation of how we construct subsamples of groups defined by mass. 
The key role played by mock catalogues of groups is also outlined 
in Section 2. The estimation of the correlation function of groups is 
set out in Section 3. A number of assumptions need to be made in 
order to measure the clustering of groups and these issues are dealt with 
in Section 4, in which we use mock catalogues extensively
to quantify the likely random and systematic errors on our measurements. 
The clustering measurements for 2dFGRS groups are presented in Section 5. 
A summary of our conclusions is given in Section 6.

\begin{table*}
\caption{\small
{
The properties of the 2PIGG samples used. A maximum redshift of $z=0.12$ 
is adopted and there are a minimum of two galaxies per group. 
The first column gives the sample label ($0$ denotes the full group sample), 
the second gives the lower luminosity limit that defines each sample, the 
third column gives the median luminosity of groups in the sample,  
the fourth column gives the mean number of galaxies per group (the value in 
brackets gives the median number of group members), the fifth column gives 
the median velocity dispersion along with half the interval containing 
$68\%$ of the groups, the sixth column gives the median mass of 
the groups (group mass is obtained using eq. 4.7 from Eke et al. 2004a), 
the seventh column gives the combined number of groups in the 
NGP and SGP regions, the eighth column gives the mean separation 
of groups and column nine gives the redshift space correlation length.
}}
\begin{tabular}{ccccccccc}
\hline
\hline
\noalign{\vglue 0.2em}
Sample & luminosity cut & median lum. & $N_{\rm member}$& $<\sigma_{g}>$ & median mass & Number &   $d_c$ & $s_0$\\ % & $\Gamma$ \\
ID 	      & $\log_{10}(L/{\rm h}^{-2}L_{\odot})$  & $ (10^{10}{\rm h}^{-2}L_{\odot})$& mean\,\,[median] & (kms$^{-1}$)  & $(10^{12}h^{-1}M_{\odot})$ & of groups &($h^{-1}$Mpc) & ($h^{-1}$Mpc)\\
\noalign{\vglue 0.2em}
\hline
\noalign{\vglue 0.2em}
 $0$ &   all    & 1.8 & $4.0\,[2]$   & $149\pm187$  & \,\,5.0 & $15938$ & \hphantom{0}$4.0$ & \hphantom{0}$5.5\pm0.1$  \\%& $-1.63$ \\
 $1$ &   $>10.2$   & 3.0 & $5.3\,[3]$   & $181\pm184$  & \,\,9.4 & \hphantom{0}$8914$  & \hphantom{0}$7.5$ & \hphantom{0}$6.0\pm0.2$  \\%& $-1.71$ \\
 $2$ &   $>10.6$  & 5.5 & $9.0\,[5]$   & $229\pm175$  & \hphantom{0}22 & \hphantom{0}$3530$  & $10.1$ & \hphantom{0}$7.8\pm0.3$  \\%& $-1.68$ \\
 $3$ &   $>10.9$   & 12 & $19\,[12]$ & $318\pm166$  & \hphantom{0}64& \hphantom{0}$1020$  & $16.2$ & $11.1\pm0.9$ \\%& $-1.91$ \\
 $4$ &   $>11.1$   & 19 & $29\,[20]$ & $378\pm196$  & 110 & \hphantom{00}$467$   & $23.4$ & $12.6\pm1.0$   \\%& $-1.86$ \\
 $5$ &   $>11.3$   & 30& $45\,[32]$ & $460\pm200$  & 200 & \hphantom{00}$211$   & $32.9$ & $16.0\pm1.7$   \\%& $-2.03$ \\
 $6$ &   $\,>11.5$  & 42 & $60\,[47]$ & $539\pm161$  & 310 & \hphantom{00}$119$   & $46.0$ & $19.4\pm2.9$   \\%& $-2.10$ \\

\noalign{\vglue 0.2em}
\hline
\hline
\end{tabular}\label{tab:subsamples}
\end{table*}

\section{The data and sample construction}
\label{sec:groups}

In this Section we describe the catalogue from which measurements 
of the clustering of groups are made, the role played by mock datasets 
in our analysis and the definition of subsamples of groups. 

We use the 2PIGG catalogue constructed by Eke et al. (2004a). 
The 2PIGG sample is the largest and most homogenous group 
catalogue currently available and therefore provides a unique tool 
with which to study the clustering of galactic systems.
The properties of the 2PIGG are outlined in Section \ref{ssec:2PIGG}. 
An important and novel feature of the Eke et al. analysis is the 
extensive use of realistic, physically motivated mock catalogues 
to calibrate the performance of the group finding algorithm and 
to understand how the recovered groups relate to the underlying 
distribution of dark matter. 
We utilize these catalogues to assess possible systematic effects in 
the measurement of clustering from the 2PIGG sample and to estimate 
the errors on our results; a brief outline of the mocks is given in 
Section \ref{ssec:mocks}.
We present the rationale behind our definition of subsamples extracted 
from the 2PIGG in Section \ref{ssec:subsamples}. 
Finally, the choice of relation between an observed group property 
and the underlying group mass is justified in Section \ref{ssec:mass}.

\subsection{The 2PIGG Catalogue}
\label{ssec:2PIGG}

The 2PIGG catalogue was constructed from the completed 2dFGRS as 
described by Eke et al. (2004a).
The group finding algorithm requires a small number of parameters 
to be set. The motivation for the adopted parameter values and 
the consequences of these choices for the accuracy and completeness of 
the catalogue are discussed at length by Eke et al. (2004a). 
Eke et al. find that after applying the identification algorithm to the
2dFGRS (which contains $\sim 190000$ galaxies in the two large 
contiguous regions), $56$ percent of the galaxies are grouped into 
$28\,877$ groups containing at least two members.  
These groups have a median velocity 
dispersion of $190 {\rm kms}^{-1}$; if groups of four or more members 
are considered, this value changes to $260{\rm kms}^{-1}$. 
In both cases, the median redshift is $z=0.11$.
The 2PIGG catalogue is sufficiently large that it may be divided into 
subsamples in order to measure trends in clustering strength with 
group properties; the construction of subsamples from the 2PIGG 
catalogue is set out in Section \ref{ssec:subsamples}.

\subsection{Mock 2PIGG catalogues} 
\label{ssec:mocks}

Two types of mock catalogue are used in this paper: semi-analytic mocks 
and Hubble Volume mocks. These mocks have different underlying physics 
and play different roles in our analysis.

\begin{itemize}
\item[(i)] {\it Semi-analytic mocks}.  We use the mock 2PIGG catalogues
constructed by Eke et al. (2004a). 
These mocks are produced from a high resolution N-body 
simulation which is populated with galaxies using the semi-analytic model, 
{\tt GALFORM} (Cole et al. 2000; Benson et al. 2002, 2003).
The simulation uses standard $\Lambda$CDM parameters, with the normalization 
of the density fluctuations given by $\sigma_{8}=0.80$ and 
a primordial spectral index of $n=0.97$, 
in line with the constraints from the first year of data from WMAP 
(Spergel et al. 2003).
The N-body simulation box is $250h^{-1}$Mpc on a side and contains 
$1.25\times10^{8}$ dark matter particles each of mass 
$1.04\times10^{10}h^{-1}M_{\odot}$. 
We use the $z=0.117$ simulation output, which is close 
to the median redshift of the 2dFGRS. 
Dark matter haloes are identified using a friends-of-friends algorithm 
with a linking length of $0.2$ times the mean interparticle 
separation (Davis et al. 1985). 

The halo resolution limit is 
taken to be $1.04\times10^{11}h^{-1}M_{\odot}$. 
The {\tt GALFORM} code is run for each halo and galaxies are 
assigned to a subset of dark matter particles in the halo 
following the technique described by Benson et al. (2000). 
Around $90\%$ of central galaxies brighter than 
$M_{b_{\rm J}}-5\log_{10} h=-17.5$ 
are expected to be in haloes resolved by the simulation. 
The effective limit of the catalogue is extended to 
$M_{b_{\rm J}}-5\log_{10} h=-16$ using a 
separate {\tt GALFORM} calculation for 
a grid of halo masses below the resolution limit of the N-body 
simulation. These galaxies are assigned to particles that are not 
identified as part of a resolved dark matter halo; such particles account 
for approximately $50\%$ of the dark matter in 
the $z=0.117$ simulation output.
The luminosity function predicted by the semi-analytic model is close 
to that estimated for the 2dFGRS (Norberg et al. 2002b): however, we 
force the luminosity function of the model galaxies to agree with the 
2dFGRS luminosity function by rescaling the luminosity of the model 
galaxies.  
A mock 2dFGRS is then extracted by placing an observer at a random 
location within the simulation cube and applying the radial and 
angular selection function of the 2dFGRS (Norberg et al. 2002b).
The simulation box needs to be replicated several times to cover the 
full 2dFGRS volume and geometry.   
For our absolute magnitude limit, the mock 2dFGRS is complete 
above $z=0.04$.
The mock 2dFGRS is then degraded by applying the redshift completeness 
mask, and the magnitude and velocity errors of the 2dFGRS 
(Norberg et al. 2002b). 
The Eke et al. group finding algorithm is applied to the mock 2dFGRS 
catalogue in exactly the same way and with the same parameter values 
as for the real data. We will henceforth refer to the resulting mock 
2PIGG catalogue as the {\tt GALFORM} mock.

The {\tt GALFORM} mock has three roles to play in our analysis: 
(1) To determine which of the observed characteristics of 2PIGG systems 
provides the most reliable estimate of the true mass of the 
underlying dark matter halo in which the group resides. 
(2) To assess the systematic errors in our clustering estimates, by 
comparing measurements from a 2PIGG mock with the measurement for an 
equivalent sample of galaxy groups extracted from the simulation cube
(i.e. before applying the 2dFGRS selection criteria and mask). 
This allows us to quantify how well our clustering estimator compensates 
for the selection function of the survey and its redshift incompleteness. 
The results of these comparisons are used to define sample selection 
criteria that, when applied to the 2PIGG catalogue, will yield the most 
robust and reliable clustering measurements.  
The construction of the equivalent samples is outlined below.
(3) To cast the predictions of the $\Lambda$CDM model for the clustering 
of galactic systems in a form that can be confronted directly with the 
observations. 

Finally, we explain the criteria used to construct samples 
of groups from the N-body simulation cube that can be compared 
directly with the groups taken from the {\tt GALFORM} mock; these 
samples will be denoted as equivalent samples. 
We will employ two approaches to construct equivalent samples.
In the first approach, the spatial abundance of groups identified
in the {\tt GALFORM} mock is reproduced in the simulation 
cube by selecting an appropriate fraction of the groups at 
each mass. 
The second method is in two stages. First, an effective bias is 
computed for the groups in the {\tt GALFORM} mock, 
using Eq. \ref{eq:beff} below, and the true mass of each group. 
Second, groups more massive than some lower mass limit in the simulation 
cube are selected, such that the effective bias of these groups matches 
that computed for the mock 2PIGG sample in the first stage. 
The correlation functions measured from these equivalent samples are 
compared with direct measurements from the {\tt GALFORM} mock in 
Section \ref{ssec:xissample}.

\vskip .3cm

\item[(ii)] {\it Hubble Volume mocks}. Eke et al. (2004a) also 
made use of mock 2dFGRS catalogues extracted from the  
$\Lambda$CDM Hubble Volume 
simulation (Jenkins et al. 2001; Evrard et al. 2002). 
In this case,  the ``galaxies'' are dark matter particles 
that are selected in order to have a clustering strength similar 
to that measured for galaxies in the flux limited 2dFGRS 
(Hawkins et al. 2003). 
The selection is based upon the smoothed density of the dark matter 
distribution (Cole et al. 1998; see also Norberg et al. 2002b).
The large volume of the Hubble Volume simulation ($27{\rm Gpc}^3$)
allows a high number of independent mock versions of the 2dFGRS to 
be extracted; the ensemble that we consider contains 22 2dFGRS mocks.
The 2PIGG group finding algorithm is run on these 
Hubble Volume mocks to produce a set of groups in each mock. 
The primary aim of these mocks is to provide an estimate of the 
error on measured clustering statistics; these errors naturally include the 
contribution from sample variance due to large-scale structure. 
We compute the {\it rms} scatter over the ensemble of 22 mocks in the 
manner described by Norberg et al. (2001). 
To recap, the correlation function measured from one of the mocks 
is considered as the ``mean'' and the scatter of the remaining 
mocks around this mean is computed. This process is repeated for 
each mock in turn. The {\it rms} scatter is the resulting mean scatter. 
This procedure gives an {\it rms} scatter that is larger than the 
formal variance over the ensemble of mocks, since, to some extent, 
it takes into account the covariance between the correlation function 
measurements in different bins.  
The fractional {\it rms} scatter obtained from the mocks is 
taken as an estimate of the statistical error and is applied to 
the measured correlation functions. 

\end{itemize}

\subsection{Subsample definition}
\label{ssec:subsamples}

One could consider dividing the 2PIGG catalogue into either cumulative 
or differential bins in a property related to group mass. 
Naively, one might anticipate that a division into differential mass bins 
would give a cleaner trend of clustering amplitude varying with increasing 
sample mass, since the clustering signal from cumulative 
samples might be dominated by the most massive objects.  
To investigate this prejudice, we use theoretical models of the 
clustering of dark matter haloes and ascertain which of these two 
alternatives is the best way to split up the 2PIGG catalogue.

The theoretical predictions for halo clustering are based upon 
the formalism developed by Cole \& Kaiser (1989)
and Mo \& White (1996; hereafter MW). 
These authors computed an asymptotic bias for dark matter haloes, 
using extended Press \& Schechter (1974) theory and the spherical 
collapse model. 
Sheth, Mo \& Tormen (2001; hereafter SMT) improved upon this calculation 
by incorporating an ellipsoidal collapse model. 
Both approaches have been tested extensively against direct predictions 
from N-body simulations (Governato et al. 1999; Colberg et al. 2000; 
SMT; Padilla \& Baugh 2002). 

\begin{figure}
{\epsfxsize=8.truecm 
\epsfbox[70 190 485 604]{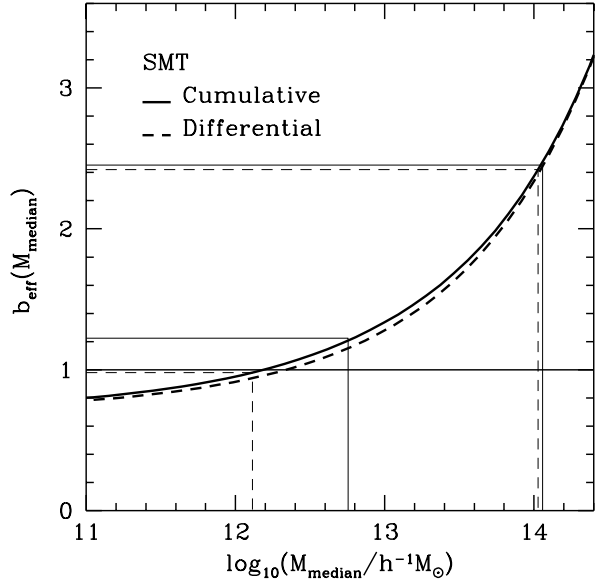}}
\caption{
Theoretical predictions for the effective bias factor, $b_{\rm eff}$, 
of a sample of dark matter haloes, as defined by Eq. \ref{eq:beff}. 
The curves show the expectations for samples of haloes defined using 
either differential (dashed) or cumulative mass bins (solid), computed 
using the formalism developed by Sheth, Mo \& Tormen (2001; SMT). 
The horizontal line indicates a bias of unity. 
The thin lines that intersect the curves and run parallel and 
perpendicular to the axes indicate the dynamic range of effective 
biases for the samples we use in this paper. The effective bias is 
computed from the {\tt GALFORM} mock and the calculation 
is described fully in Section \ref{ssec:mass}. The thin dashed lines show 
the predictions for $2$ or more members per group and the thin solid lines 
show the results for a minimum of $4$ members per group. 
}
\label{fig:bias}
\end{figure}

The effective bias for a sample of dark matter haloes is computed 
by weighting the bias factor associated with each individual halo, 
$b(M)$,  
by its abundance in the sample (e.g. Baugh et al. 1999; Padilla \& 
Baugh 2002):
\begin{equation}
b_{{\rm eff}}=\frac{ 
\int_{M_{1}}^{M_{2}} b(M) \left({\rm d}n(M)/{\rm d}M\right) {\rm d}M 
}
{ \int_{M_{1}}^{M_{2}} \left({\rm d}n(M)/{\rm d}M\right) {\rm d}M },
\label{eq:beff}
\end{equation}
where ${\rm d}n(M)/{\rm d}M$ is the space density of haloes in the 
mass interval $M$ to $M+\delta M$.
In the case of cumulative mass bins, the lower mass limit, $M_{1}$, 
defines the sample, as $M_{2}\rightarrow \infty$.

Fig. \ref{fig:bias} shows the theoretical predictions for the 
effective bias as a function of the median mass of the sample. 
For differential mass samples, an interval of one decade in mass 
is assumed.
To make these predictions, we have adopted the linear theory 
power spectrum used in the N-body simulation described in Section 
\ref{ssec:mocks}.
The clustering amplitude of the overall dark matter distribution 
corresponds to $b=1$.  
The MW theory predicts that haloes with mass below a characteristic 
value $M_{*}$ (defined as the mass contained within a sphere for which 
the linear theory variance is equal to the threshold for collapse) 
will display a weaker clustering signal than the overall mass 
distribution, i.e. these haloes will have $b(M)<1$. 

The dynamic range in mass of the 2PIGG samples that we consider is 
$\approx 10^{12} $h$^{-1}M_{\odot}$  to  $10^{14} $h$^{-1}M_{\odot}$.  
Over this interval, the theoretical predictions for cumulative and 
differential mass bins have similar shapes, so there is no clear 
advantage in using differential mass bins. 
Moreover, the cumulative samples benefit from better statistics. 
For these reasons, we will subdivide the 2PIGG sample using cumulative 
bins in a group property related to mass (see Section \ref{ssec:mass}).

\subsection{Indicator of group mass}
\label{ssec:mass}

\begin{figure}
{\epsfxsize=8.5truecm 
\epsfbox[10 168 574 698]{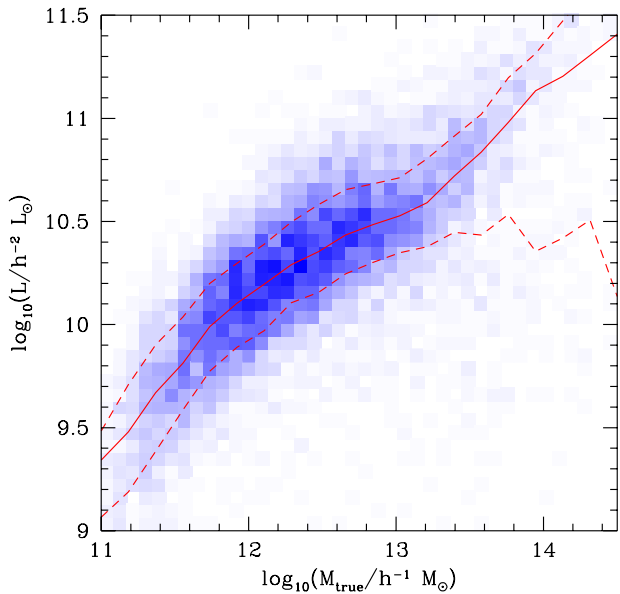}}
\caption{
The corrected total group luminosity plotted as a function of true mass 
for groups extracted from the {\tt GALFORM} mock with $z < 0.12$.
The grey scale indicates the number of groups,  with a darker shading 
indicating a higher number of groups. The solid line shows the median, 
and the dashed lines enclose the interval within which $68\%$ of the 
groups lie.
}
\label{fig:ml}
\end{figure}

Our goal in this paper is to estimate the clustering of subsamples 
of the 2PIGG catalogue, defined by group mass. 
We therefore need to find an observed property of the 2PIGGs 
that displays a tight correlation with the underlying or ``true'' group mass. 
Following Eke et al. (2004a), we have considered two possibilities: 
(i) A dynamical mass estimate that depends upon the {\it rms} radius of 
the group and its estimated velocity dispersion. 
(ii) An indirect estimate in which the corrected total 
group luminosity is used to infer a mass from the estimated 
median mass to light ratio of 2PIGGs.

From the {\tt GALFORM} mock catalogue, it is possible to find 
the ``true" mass of any group by associating it with a dark matter halo 
in the simulation box. The group's ``true" mass is simply the mass of the 
halo, as computed by summing the particles identified as members of the 
halo by a friends-of-friends group finding algorithm.  
Fig. \ref{fig:ml} shows the relation between corrected total group luminosity 
and the true group mass for the {\tt GALFORM} mock.  
The total group luminosity is obtained by summing the luminosity of the 
galaxies identified as members of each group, and then applying a 
correction to account for missing galaxies that are 
fainter than the apparent magnitude limit of the 2dFGRS. 
This correction factor is redshift dependent and reaches a 
factor of $\sim 2$ at $z\sim0.12$. 
In Fig. \ref{fig:ml}, we only show groups with $z<0.12$.
The shading represents the abundance of
groups for a given total luminosity and true mass; the darker shading 
indicates a higher number density of groups.
The solid line shows the median true mass to total luminosity relation, 
and the dashed lines enclose $68\%$ of the groups in each mass bin.
This relation is reasonably tight over more than 
two decades in true mass. The equivalent comparison between the 
dynamical mass estimate and the true mass exhibits much more 
scatter (see Eke et al. 2004b).
The dynamical mass estimator is a particularly poor indicator of 
true mass for objects less massive than $10^{13}h^{-1}M_{\odot}$.
We therefore choose to employ the corrected total group luminosity as the  
indicator of the true group mass.  
However, caution is advisable when dealing with very massive groups, 
since, as is apparent in Fig. \ref{fig:ml}, the interval containing 
$68\%$ of the groups with masses $M>10^{14}h^{-1}M_{\odot}$ 
broadens considerably.
This is due to the fact that such large groups, because of their rarity, 
are more likely to be found at higher redshifts, and therefore require a 
larger correction to be applied to the observed group luminosity  
to infer the total group luminosity.

\begin{figure}
{\epsfxsize=8.5truecm 
\epsfbox[63 188 574 664]{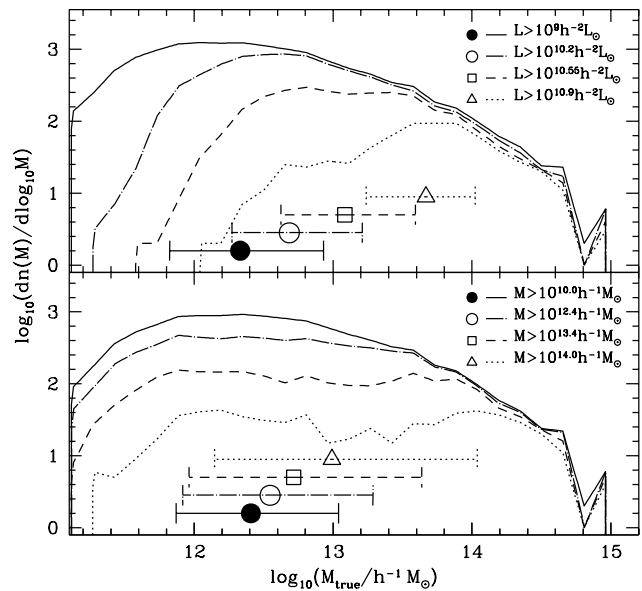}}
\caption{
The distribution of true mass for subsamples of mock 2PIGG groups with 
at least $2$ galaxy members.
The upper panel shows the mass distributions for subsamples selected 
by cumulative total luminosity and the lower panel shows 
the results for samples selected by 
cumulative bins in dynamical mass; the bins are given 
by the key in each panel. 
The points with error bars show the median and the 10-80 percentile 
range in true mass for each bin. 
}
\label{fig:massdist}
\end{figure}

Further evidence supporting our choice of the total group luminosity as a mass
indicator is given in Fig. \ref{fig:massdist}.  Here, we show the ``true" 
mass distributions of samples from the {\tt GALFORM} mock selected according
to two criteria: (i) cumulative bins in total luminosity (top panel), 
and (ii) cumulative bins in dynamical mass (lower panel). The sample 
definitions are given by the key in each panel. 
These distributions are for groups with at least $2$ members.
The different lines in each panel show the corresponding 
distributions of true group mass for each subsample. 
The points with error bars show the median and the 
range containing $60\%$ of the 
groups for the true mass distributions.
It is clear from this plot that the distribution of true mass 
for samples selected by total group luminosity spans a larger dynamic range 
in mass, and the distributions have tighter percentile intervals.  
We have performed the same exercise for groups with a minimum of four members 
and reach similar conclusions. 
However, we note that in the case of groups with 4 or more members, 
there are fewer low mass groups as expected.

We now investigate how the clustering signal from group samples 
varies with the minimum number of group members, $n_{\rm min}$, 
looking at the cases where $n_{\rm min}=2$ or $4$.
We compute an effective bias for each sample using Eq. \ref{eq:beff}.
In Fig. \ref{fig:bias}, the effective bias is plotted at a 
representative mass for each sample 
(indicated by the lines which intersect the bias curves and run 
parallel and perpendicular to the mass axis). This mass is obtained 
using the {\tt GALFORM} mocks by finding the halo mass above which the 
effective bias %space density 
of haloes in the simulation cube matches that of the  
haloes recovered from the mock group catalogue. 
We show the effective bias for the brightest and faintest 
cumulative luminosity bins to illustrate the dynamic range of 
our clustering measurements. 
The dashed lines show the results of this calculation for $n_{\rm min}=2$ and 
the solid lines show the case where $n_{\rm min}=4$. 
The range of effective masses (x-axis) and clustering strengths (y-axis) 
probed by the different samples is widest when considering groups 
with $n_{\rm min}=2$.

\section{Measuring the clustering of groups} 
\label{sec:method}

We now outline the method followed to measure the clustering of 
the 2PIGG samples. In a sample of groups extracted from a flux limited galaxy 
redshift survey, the space density of groups is a function 
of radial distance. An accurate estimate of the radial selection 
function of the sample is required to make a robust measurement of the 
clustering of groups. 
The procedure that we follow to obtain the selection function 
from the redshift distribution of groups is set out in  
Section \ref{ssec:nofz}.
The weighting scheme used to approximate a minimum variance estimate 
is given in Section \ref{ssec:xi}, along with the estimator used to 
infer the correlation function. 

\subsection{The radial selection function} 
\label{ssec:nofz}

An accurate estimate of the selection function of groups in 
the 2PIGG samples is required to permit an analysis of the 
clustering signal of the groups. 

The most straightforward way of doing this is to use the luminosity 
function of groups to obtain an estimate of the selection function 
which is unaffected by large scale structure. 
However, Eke et al. (2004c) show, using the {\tt GALFORM} mock, 
that estimates of the group luminosity function are affected 
by errors in the determination of total group luminosity and 
by contamination, leading to systematic errors in the number 
density of groups of up to a factor of $4$.  

\begin{figure}
{\epsfxsize=8.truecm 
\epsfbox[20 178 450 702]{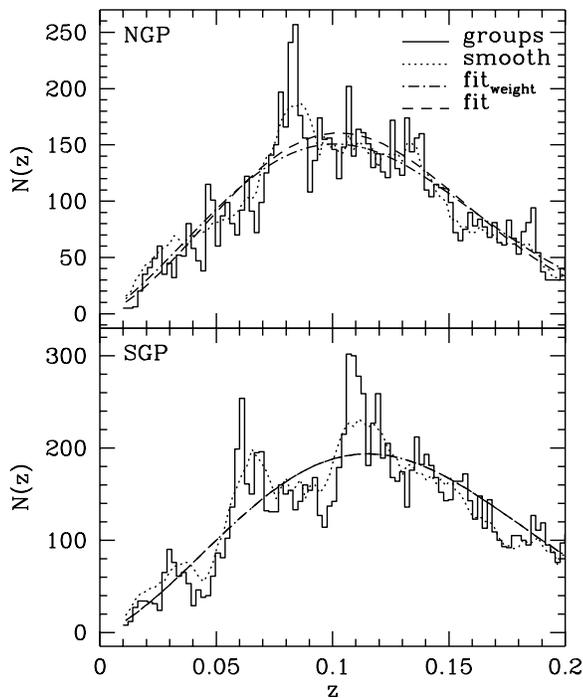}}
\caption{
The redshift distributions of the 2PIGG catalogue (histograms);
the upper panel corresponds to the NGP region and the lower
to the SGP region.  The redshift distribution after smoothing is 
shown by the dotted line (see text for details). 
The dot-dashed lines show volume weighted fits (Eqs. \ref{eq:fitnz} and 
\ref{eq:chiv}), and the dashed lines show unweighted fits  (Eqs.
\ref{eq:fitnz} and \ref{eq:chi}). 
}
\label{fig:nz}
\end{figure}

The alternative is to estimate the selection function directly from 
the observed redshift distribution. The concern in this case 
is that the redshift distribution displays features that are 
due to large scale structure (see the histogram in Fig. 
\ref{fig:nz}). Care must be taken when fitting a parametric 
form to the observed redshift distribution to avoid `over-fitting' 
the peaks and troughs, thereby inadvertently removing some of 
the clustering signal from the estimated correlation function. 

With this concern in mind, we explore two procedures to describe 
the shape of the observed redshift distribution: fitting 
parametric forms or smoothing the observed redshift distribution.
We will explore the influence on the measured correlation 
function of these different ways of 
fitting the redshift distribution in Section \ref{ssec:optimal}.

We adopt a fit to $N_{\rm obs}(z)$, the observed 2PIGG 
redshift distribution, of the form
\begin{equation}
N_{\rm fit}(z){\rm d}z=z^{a-1} \times \exp \left( \frac{z}{b} \right)^a{\rm d}z,
\label{eq:fitnz}
\end{equation}
where $a$ and $b$ are parameters set by minimising the quantity
\begin{equation}
\chi_{N(z)}^2=\sum_{i=1}^{\rm nbins} \frac{(N_{\rm obs}(z_i)-N_{\rm fit}(z_i))^2}{N_{\rm obs}^{2}(z_i)}.
\label{eq:chi}
\end{equation}
Alternatively, we have experimented with including 
a weight proportional to the volume of each redshift 
bin of width $\Delta z$,
\begin{equation}
\chi_{v,N(z)}^2=\sum_{i=1}^{\rm nbins} \frac{(N_{\rm obs}(z_i)-N_{\rm fit}(z_i))^2}{N_{\rm obs}^{2}(z_i)}z_i^2\Delta z_i.
\label{eq:chiv}
\end{equation}
The effect on the best fit parameter values of weighting by 
volume is small. 
From now on, best fit parameter values $a$ and $b$ obtained using Eq. 
\ref{eq:chi} (and the redshift distribution constructed using them) 
will be referred to as unweighted fits, while those found using 
Eq. \ref{eq:chiv} will be referred to as volume-weighted fits.
The resulting best fits for the full 2PIGG catalogues are shown in 
Fig. \ref{fig:nz}, as indicated by the key in the upper panel.

The parametric form adopted for the fit is particularly attractive  
since it can be integrated analytically to give the cumulative 
redshift distribution 
\begin{equation}
\int_0^z N_{\rm fit}(z) {\rm d}z \big/ \int_0^{\infty} N_{\rm fit}(z){\rm d}z 
= 1-\exp \left( \frac{z}{b} \right)^a.
\label{eq:cumnz}
\end{equation}

The smoothed redshift distribution is obtained by smoothing the observed 
redshift distribution, which is tabulated in bins of width $\Delta z=0.0025$, 
with a top-hat window in redshift. 
The smoothed redshift distribution in Fig. \ref{fig:nz} was produced 
by smoothing the observed distribution with a top-hat of width 
$\Delta z = 0.01$. 

\subsection{Estimating the correlation function} 
\label{ssec:xi}

To obtain a minimum variance estimate of the correlation function,  
each group is assigned a weight given by (Efstathiou 1988): 
\begin{equation}
w_i = \frac1{1+4\pi n(z_i) J_{3}(s) c},
\label{eq:w}
\end{equation}
where we assume $J_3(s)=1200$h$^{-3}$Mpc$^3$, $c$ is the angular 
completeness of the 2dFGRS and $n(z_{i})$ is the space density of 
groups at redshift $z_{i}$.
In Section \ref{ssec:optimal} 
we explore the impact of the choice of $J_3$ on the measured 
correlation function. 
The space density of groups, $n(z)$, is obtained by integrating 
over the product of the cosmological volume element and 
the adopted description of the redshift distribution;
as explained above, this 
could be a smoothed version of the redshift distribution or a 
parametric fit.  
A catalogue of unclustered points is produced with the same radial 
and angular selection functions as the data. 
The number of random points is set to be $10$ times the 
number of data points, with a lower limit of $100000$.
In the case of the unclustered catalogue, the weights are scaled in
order to ensure that their sum matches that of the weights of the 
observed groups: 
\begin{equation}
w_{R,i}=\frac{w_{R,i}}{\sum_j w_{R,j}} \sum_j w_{D,j},
\end{equation}
where $R$ and $D$ indicate unclustered points and data groups respectively.

We adopt the Landy \& Szalay (1993) estimator for calculating
the redshift-space correlation function,
\begin{equation}
\xi(s)=(DD(s)-2DR(s)+RR(s))/RR(s),
\end{equation}
where $$XY(s)=\sum_{ij} w_{X,i} w_{Y,j},$$
with $X$ and $Y$ representing the actual groups ($D$) and/or
points in the random catalogue ($R$); $i$ and $j$ are the 
indices for each individual group/random point, and $w$ 
corresponds to the weight defined above.  

We use the same estimator to obtain the correlation function as 
a function of pair separation perpendicular ($\sigma$) 
and parallel ($\pi$) to the line-of-sight, $\xi(\sigma,\pi)$.  
From this correlation function, we can define the projected 
correlation function
\begin{equation}
\frac{\Xi(\sigma)}{\sigma}=\frac2{\sigma} \int_0^{\infty}\xi(\sigma,\pi) \rm{d}\sigma.
\label{eq:si}
\end{equation}
This quantity is free from redshift-space effects 
arising from peculiar motions and can be related directly to 
the real space correlation function (e.g. Norberg et al. 2001).

\section{Testing the estimation of the correlation function}
\begin{figure*}
\begin{picture}(450,580)
\put(0,275){\psfig{file=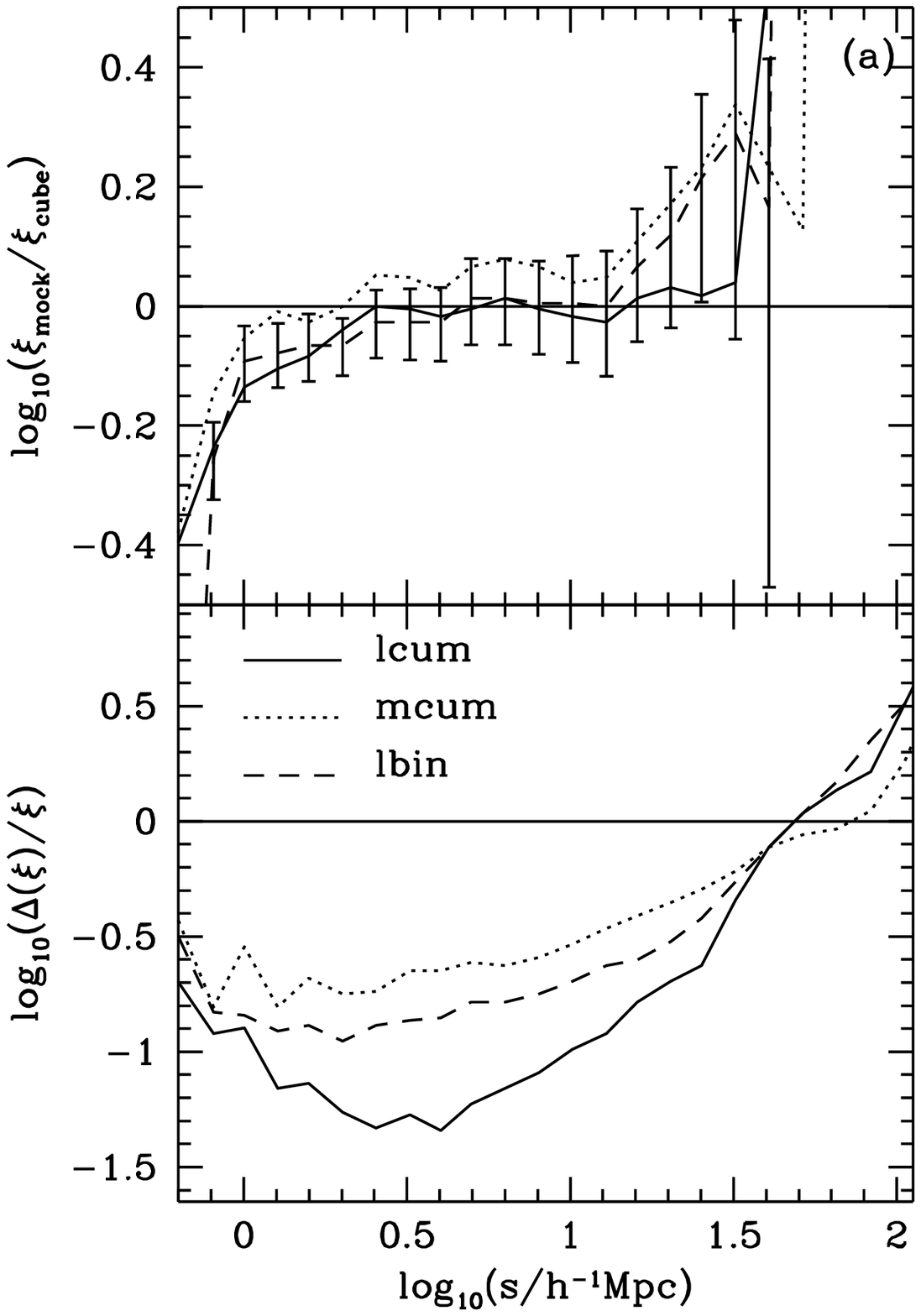,width=11cm}}
\put(245,275){\psfig{file=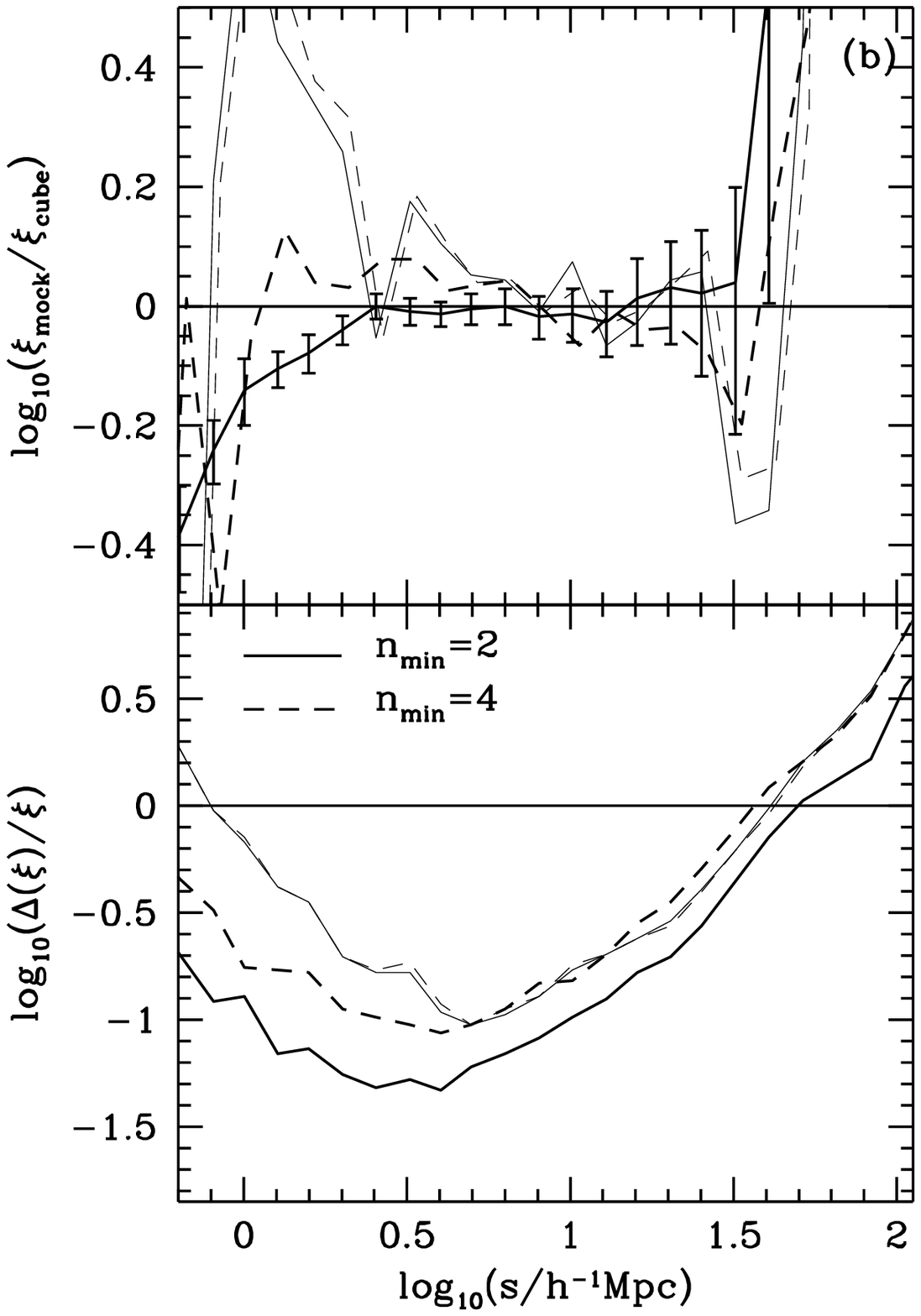,width=11cm}}
\put(0,-10){\psfig{file=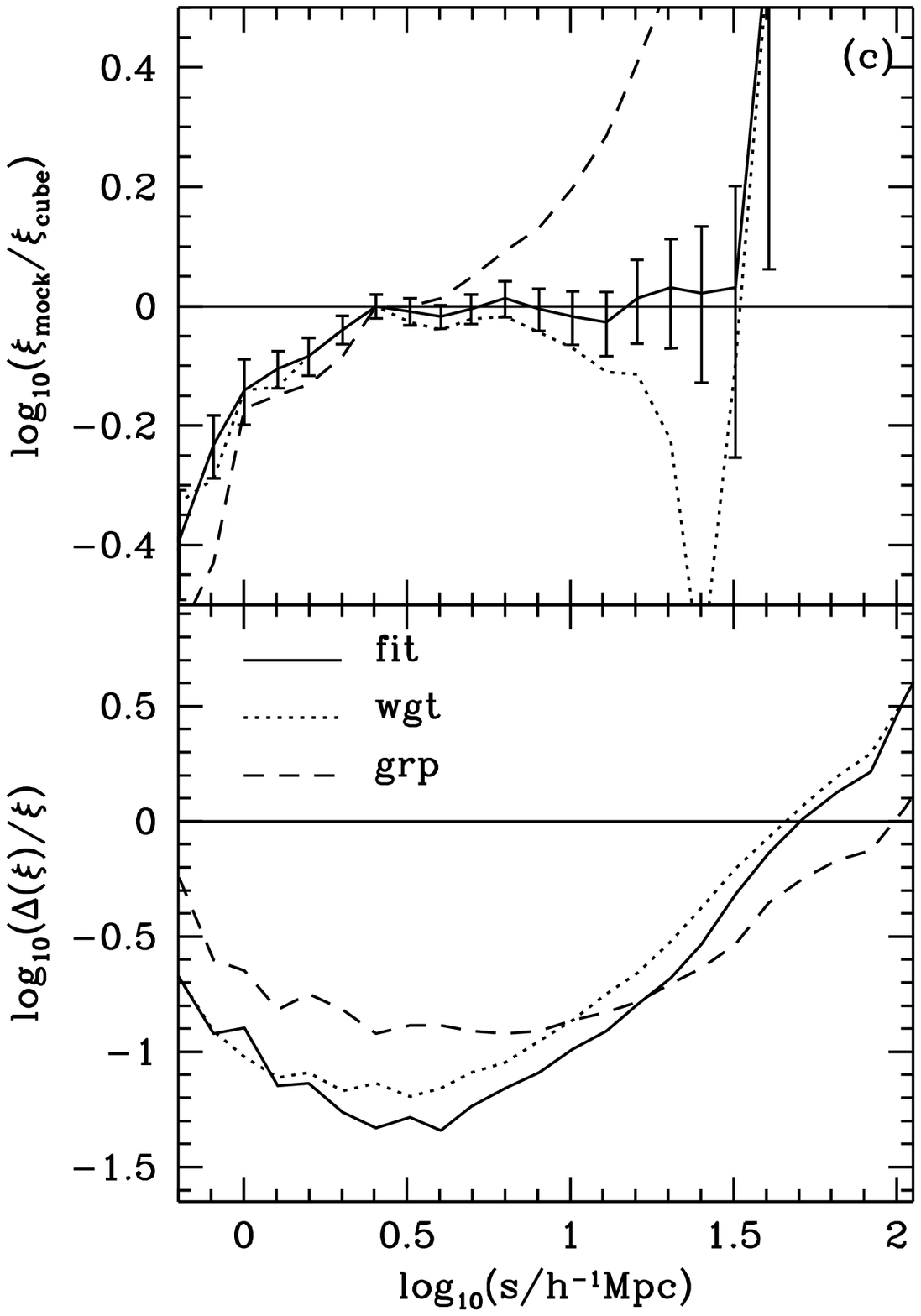,width=11cm}}
\put(245,-10){\psfig{file=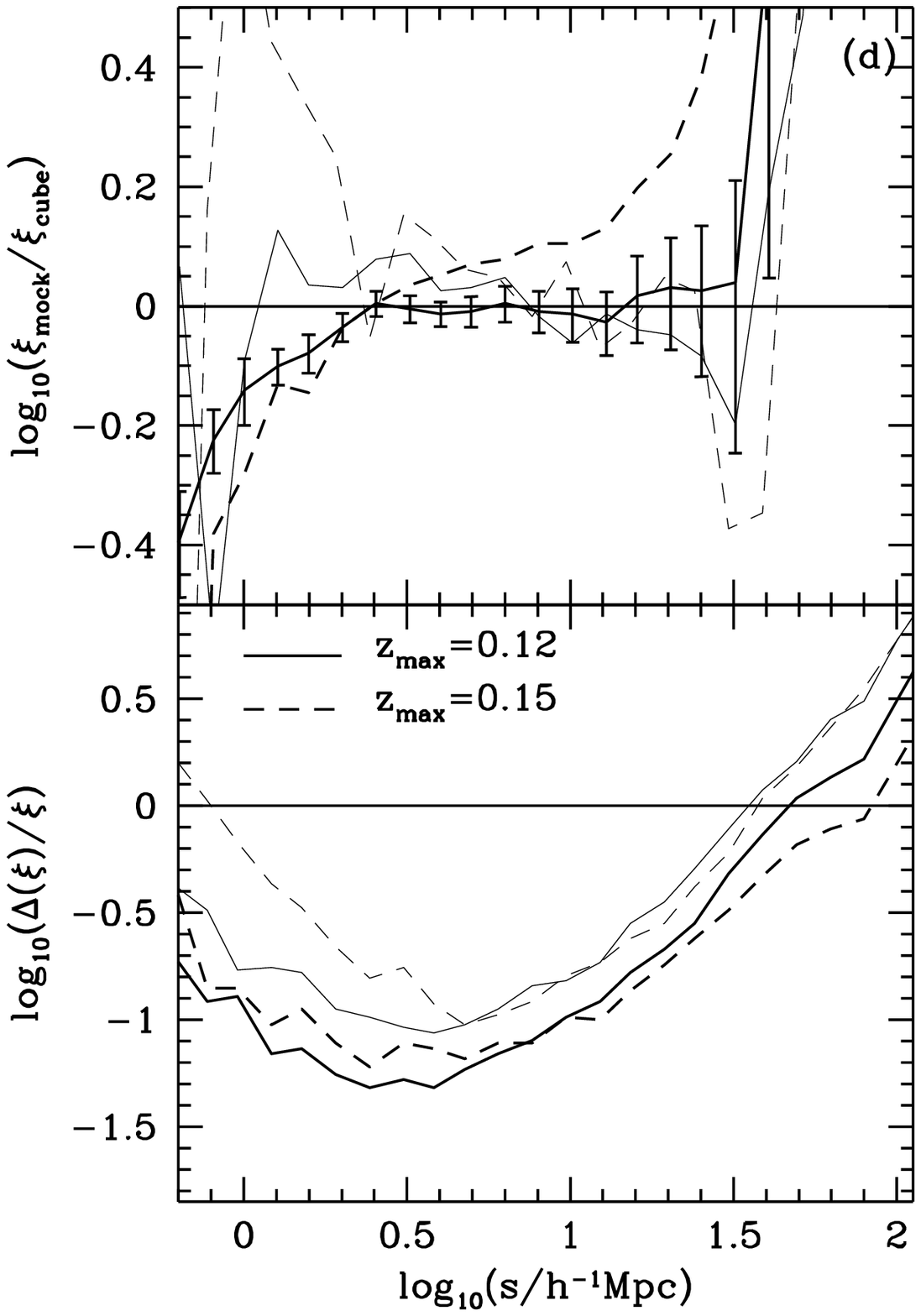,width=11cm}}
\end{picture}
\caption{
The accuracy with which the correlation function can be measured from 
the {\tt GALFORM} mock catalogue for different assumptions 
and approximations. In each 
plot, the upper panel shows the systematic error on the measured correlation 
function, expressed as the ratio of the correlation function measured from 
a mock 2PIGG sample to that measured from an idealized, comparable sample. 
The errors on this ratio come from the {\it rms} scatter over the 
ensemble of Hubble Volume mocks; these are used to plot the logarithm of 
the relative errors in the lower panel of each plot. 
The different plots show how well the correlation function can be measured 
when the following are varied:  
(a) The sample definition, as given 
in the text: the solid lines shows results for a sample 
of groups brighter than a minimum luminosity, the dotted line shows 
a sample of groups more massive than a minimum dynamical mass limit 
and the dashed line shows the case of groups within a 
differential luminosity bin.
(b) The minimum group membership. 
The thinner lines show results for a brighter sample of groups ($L>10^{10.9}
$h$^{-2}$ L$_{\odot}$. 
(c) The description of the redshift distribution used to model 
the selection function of groups. 
(d) The maximum redshift used to define the catalogue; again the thinner  
lines show results for a brighter sample.  
}
\label{fig:err}
\end{figure*}

The measurement of the clustering of groups requires a number 
of choices to be made that can have an impact upon the 
estimated correlation function. 
In this section, we carry out a careful examination of our 
method for measuring the clustering of the mock groups, with the 
goal of determining a correlation function that is accurate, 
i.e. with minimal systematic errors, and with random measurement 
errors that are as small as possible. 
The mock catalogues described in Section \ref{ssec:mocks} 
play a key role in this exercise. 
The {\tt GALFORM} mock is used to assess systematic errors by 
comparing the correlation functions measured from the semi-analytic 
mock group catalogue with the correlation function of an equivalent 
sample in the simulation cube (the construction of such  
a catalogue is described in Section \ref{ssec:mocks}). 
We stress that the goal of quantifying the systematic errors 
is to devise a clustering measurement algorithm that minimizes these errors 
rather than to correct for them.
The Hubble Volume mocks are used to estimate the random measurement 
errors, which automatically include the contribution of sample 
variance from large scale structure. 

An important test for systematic errors in our clustering 
measurement is to compare the correlation function recovered 
from a mock group catalogue with that of a comparable, equivalent 
sample drawn from the distribution of groups in the simulation cube.

The rest of this section is split into two parts. In the first, we 
present a systematic study of how various assumptions 
and approximations affect the recovered correlation function. 
We focus attention on the size of the random measurement errors 
and on minimizing systematic errors in the estimated correlation 
function. In the second part of this section, 
we build on the conclusions of the first part and apply 
an optimal method to measure the clustering of mock 2PIGGs to 
determine how well we can recover trends in clustering 
amplitude for different subsamples of groups. 

\subsection{An optimal measurement of clustering}
\label{ssec:optimal}

The main results of this subsection are presented in Fig. \ref{fig:err}. 
Each plot in this figure is divided into two panels. The top panel in 
all cases compares the correlation function recovered from a {\tt GALFORM} 
mock group catalogue with the correlation function of an equivalent 
sample, drawn from the simulation cube, of groups generated by {\tt GALFORM}. 
The errors on this ratio are the {\it rms} scatter over the ensemble of 
22 Hubble Volume group mocks.  Note that this estimate of errors includes
cosmic variance, and therefore overestimates the error bars in a
comparison between results
from the mock and the simulation cube, since both come from the same
simulated volume. In the absence of systematic errors in the 
clustering measurement, this ratio would be consistent with unity, indicated 
by the horizontal line. 
The lower panel in each plot shows the logarithm of the relative error 
on the measured correlation function, as deduced from the {\it rms} 
scatter over the Hubble Volume mocks. The horizontal line indicates an 
error of $100\%$. To improve the clarity of the presentation, we have 
used a power law fit to the correlation function for separations larger 
than $8h^{-1}{\rm Mpc}$ in the definition of the relative error, 
which avoids dividing by small or negative numbers. The correlation 
function of the equivalent sample of groups falls below 
a power law on large scales, typically $s\gsim15 h^{-1}{\rm Mpc}$, so 
the plotted error is a lower limit to the true relative error on these
scales. 
We now go through the set of choices or assumptions that needs to be made 
in our clustering measurement method, directing the reader to 
Fig. \ref{fig:err} where appropriate.

(i) {\it Choice of 2dFGRS region.} 
The differences between the correlation functions measured in 
the NGP and SGP regions provide a crude estimate of the errors; 
these differences are comparable to the errors inferred from the 
Hubble Volume mocks.  
We find that the correlation functions measured in mock 
NGP and SGP 2dFGRS regions are consistent within the errors. 
Combining the pair counts of groups in the two regions reduces 
the Poisson noise and sample variance in the clustering measurement. 
Henceforth, our results will be for combined NGP and SGP samples.

(ii) {\it Choice of $J_3$ in the radial weight.} 
The weight assigned to each group to compensate for the radial selection 
function requires a value for the integral of the two-point correlation 
function, $J_3(s)$ (Eq. \ref{eq:w}).
We have explored using constant values in the range 
$400 < J_3/$h$^{-3}$Mpc$^3 < 2400$, 
and find that the changes in the estimated correlation function 
are small and within the statistical errors. 
We adopt $J_3=1200$h$^{-3}$Mpc$^3$, which corresponds to the typical 
asymptotic value of $J_3(s)$ for separations $s>10$h$^{-1}$Mpc in
the Hubble Volume mocks.

(iii) {\it Sample definition.}
Fig. \ref{fig:err}(a) compares the correlation function measured for 
samples defined in different ways: by setting (i) a lower limit on 
total group luminosity, $L>10^{10.2}$h$^{-2}L_{\odot}$ (solid line), 
(ii) a lower limit on dynamical mass, 
$M_{\rm dyn}>10^{12.4}h^{-1}M_{\odot}$ (dotted line), 
which, for the median 
mass-to-light ratio of the 2PIGGs (Eke et al. 2004b), 
corresponds to the luminosity cut in (i), and (iii) a differential luminosity 
sample defined by $10^{10.25}<L/$h$^{-2}L_{\odot}<10^{10.55}$ (dashed line). 
The correlation function measured for the sample of groups brighter 
than the luminosity threshold in (i) has the smallest systematic 
and random errors. 
This adds further weight to the preference arrived at for cumulative 
bins in luminosity in Section \ref{ssec:mass}. 
The three remaining panels, Fig. \ref{fig:err}(b),(c) and (d) show the errors 
on $\xi(s)$ for samples of mock groups with $L>10^{10.2}$h$^{-2}L_{\odot}$ 
(shown by solid lines), and, in some cases for a brighter sample with  
$L>10^{10.9}h^{-1}M_{\odot}$ (indicated by dashed lines).

(iv) {\it Minimum number of group members}. 
We compare the accuracy of the correlation function for 
groups with a minimum number of members of $n_{\rm min}=2$ (solid  
lines) and $n_{\rm min}=4$ (dashed lines) in Fig. \ref{fig:err}(b).
The systematic errors on the correlation functions recovered from 
the mocks are roughly comparable in the two cases. 
The random errors, however, are smaller for the case of $n_{\rm min}=2$.
This is expected since this sample contains more groups. 
We recall that samples of groups with $n_{\rm min}=2$ span 
a wider range in mass (see Fig \ref{fig:massdist} 
in Section \ref{ssec:mass}).  
The dashed lines in Fig. \ref{fig:err}(b) show the results for a brighter 
sample of groups, as explained at the end of (iii) above. This sample 
contains fewer groups and so the random errors are somewhat larger than 
before, particularly at small separations. The systematic errors in the 
correlation function recovered from the mock 2PIGG also become significant 
below $s \approx 2 h^{-1}$Mpc for the brighter sample.

\begin{figure}
{\epsfxsize=8.truecm 
\epsfbox[20 170 571 700]{bwfigs/err1200_datlcum_0.12_nmin2_n_0.2.ps}}
\caption{
The relative errors on our measurement of $\xi(s)$, 
obtained by dividing the {\it rms} scatter in $\xi(s)$ from the Hubble Volume 
mocks by the mean value of $\xi(s)$ for $s<8$h$^{-1}$Mpc, and
a fit to the mean $\xi(s)$ (over
$8<s/($h$^{-1}$Mpc$)<15$) for $s>8$h$^{-1}$Mpc.  
The horizontal black line shows the $100\%$ error limit. Different 
line types correspond to different subsamples as indicated by 
the key.  
}
\label{fig:xierr}
\end{figure}

(v) {\it Selection function model.}
The impact on the correlation function of using different approximations 
to the form of the redshift distribution of groups to model 
their selection function is shown in Fig. \ref{fig:err}(c). 
The solid line shows how well the 
correlation function can be measured using a parametric fit to the 
redshift distribution of groups in the mock, as described in Section 
\ref{ssec:nofz}; the dotted line shows the results when a volume 
weighting is applied to determine the best fit parameters. 
On small scales, $s<3h^{-1}$Mpc, the systematic and random errors are 
very similar in these two cases. On larger scales however, the unweighted 
fit results in a more faithful and less noisy measurement of the 
correlation function.  
The evidence in this panel shows that the approach of using the smoothed 
redshift distribution of groups to model their selection function (dashed 
line)
is clearly flawed, leading to a significantly discrepant 
estimate of the correlation function for separations $s>1h^{-1}$Mpc. 

\begin{figure*}
\begin{picture}(450,450)
\put(0,225){\psfig{file=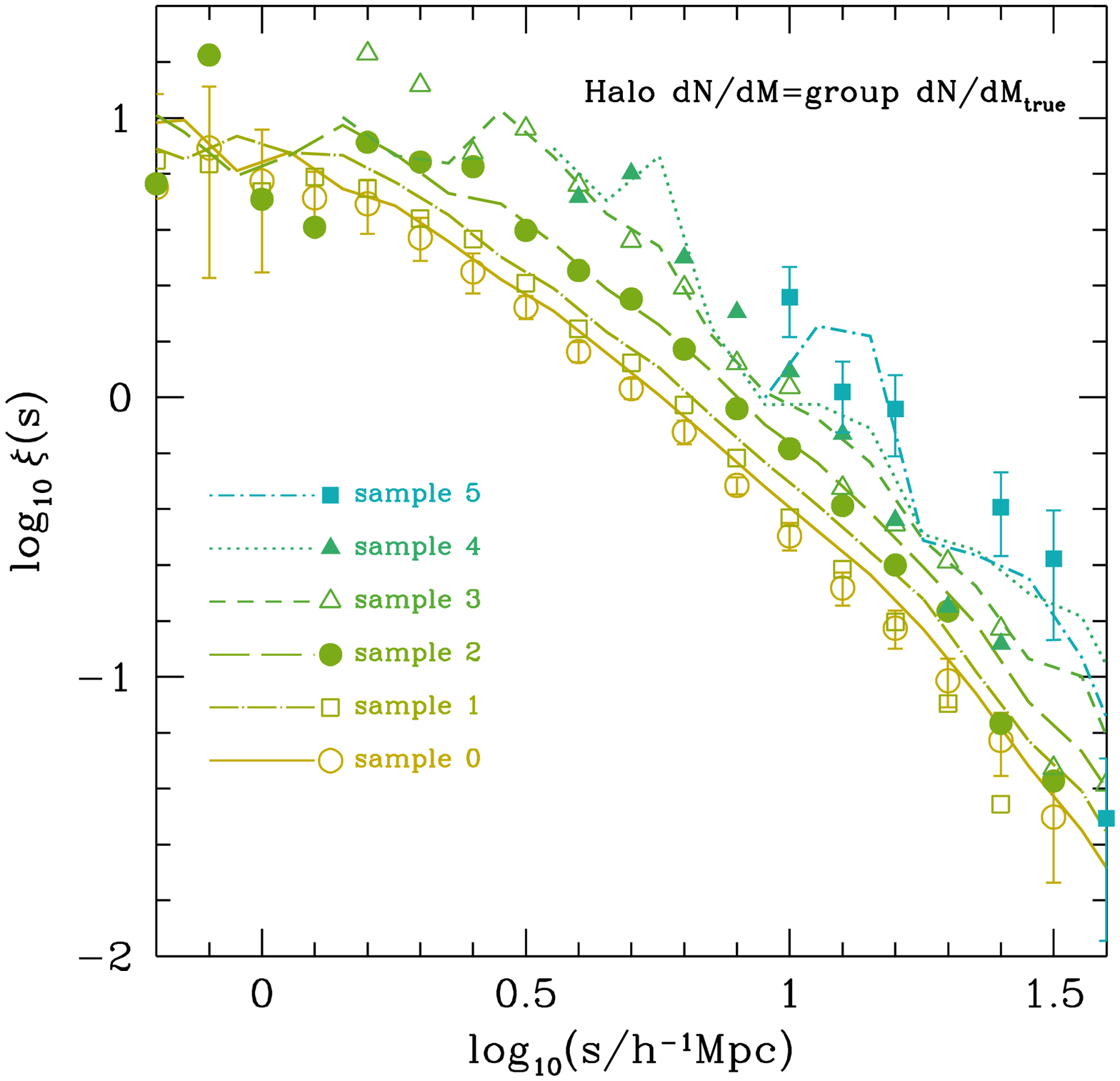,width=8cm,height=8cm}}
\put(0,0){\psfig{file=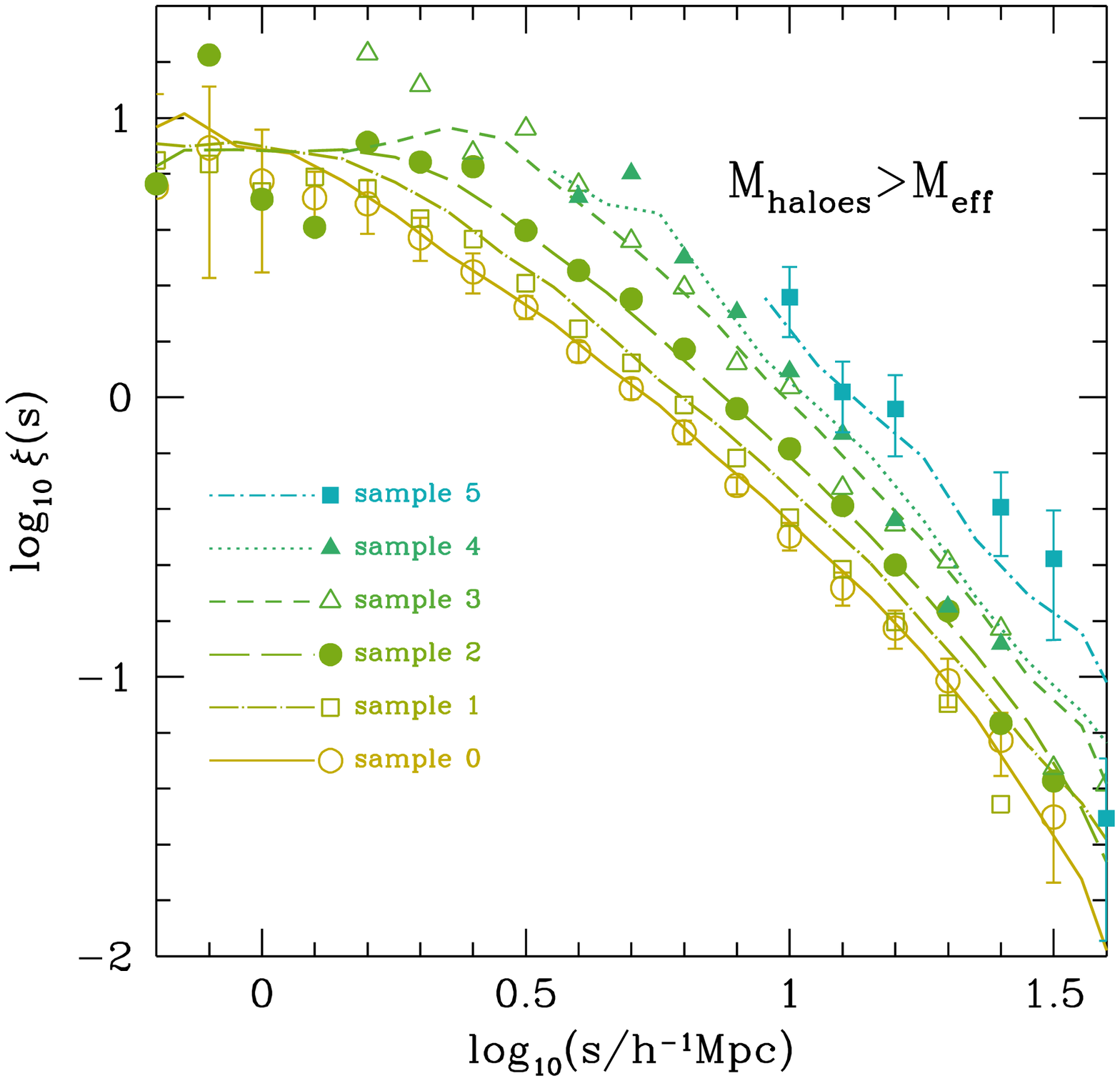,width=8cm,height=8cm}}
\put(225,225){\psfig{file=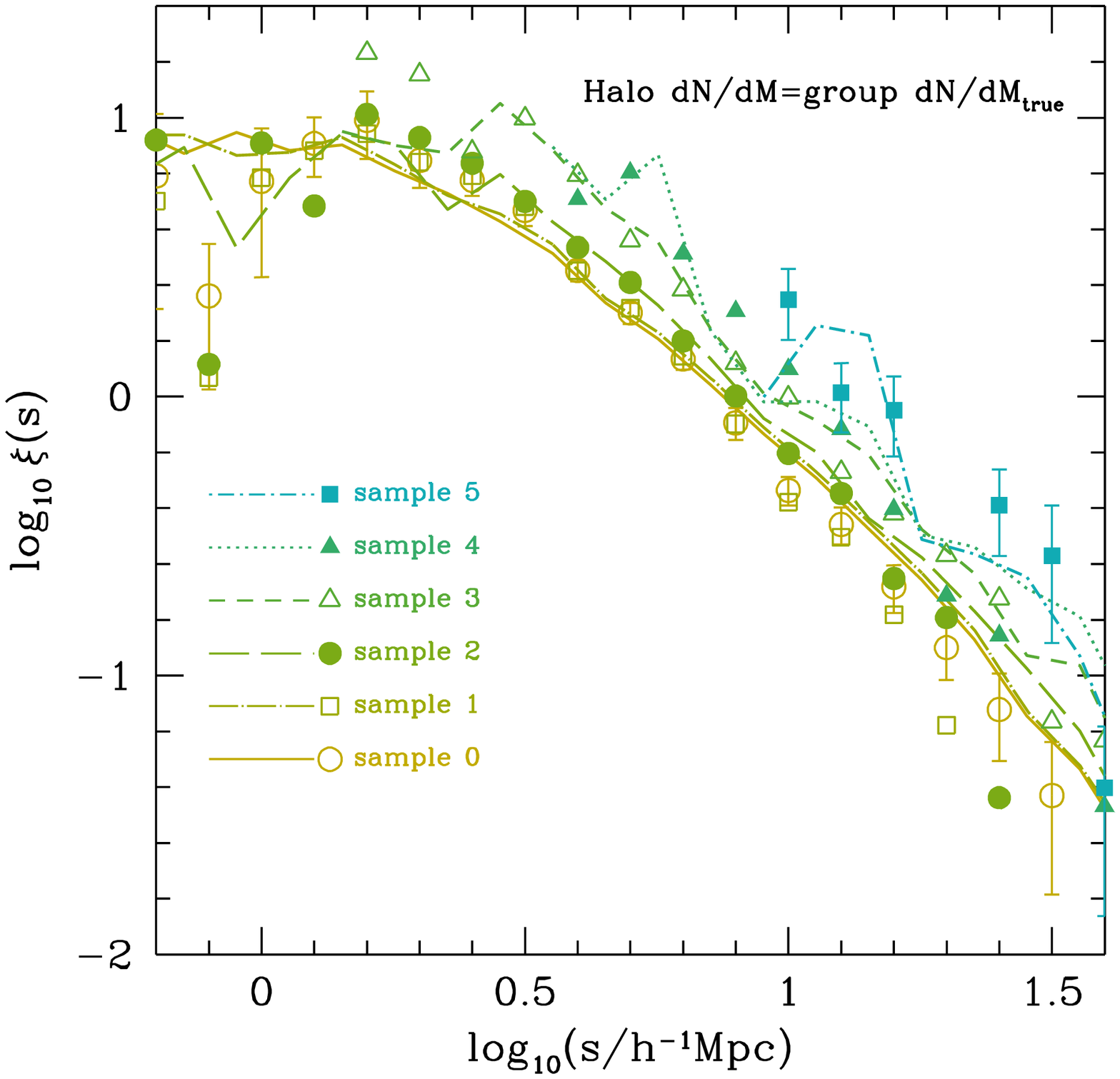,width=8cm,height=8cm}}
\put(225,0){\psfig{file=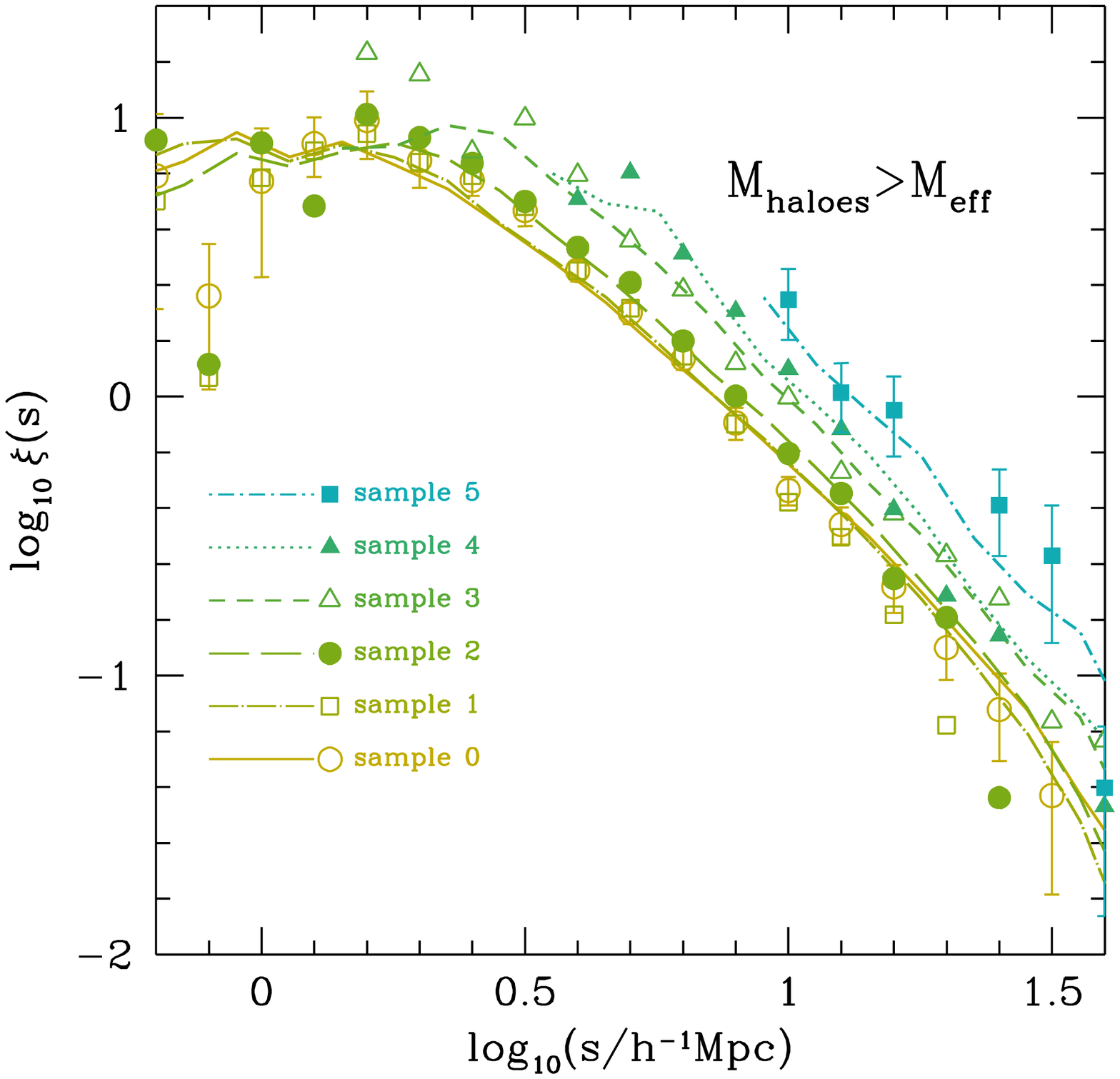,width=8cm,height=8cm}}
\end{picture}
\caption{
The correlation functions measured in the {\tt GALFORM} 2PIGG mock for 
samples defined by lower group luminosity limits (symbols with error bars).
In the left hand column, results are shown for $n_{\rm min}=2$ and in 
the right for $n_{\rm min}=4$. 
These measurements are compared with the correlation functions of equivalent 
samples of groups taken from the simulation cube (lines). 
Two different methods of constructing these equivalent samples have 
been used. In the top row, the equivalent samples have the same mass function 
as the groups in the sample from the mock 2PIGG catalogue. In the bottom row, 
the equivalent samples are complete above some halo mass, chosen so that the 
groups have the same effective bias as those recovered from the mock 2PIGG 
catalogue.
}
\label{fig:ximockcube}
\end{figure*}

(vi) {\it Sample redshift limit.}
An increase in the redshift limit of the sample leads to a larger 
volume and the associated increase in the size of the sample. 
However, these additional groups 
come at the expense of requiring a larger correction to the 
observed group luminosity to deduce the total group luminosity. 
The consequences of varying the maximum redshift limit of the sample 
are presented in Fig. \ref{fig:err}(d), where the cases 
$z_{\rm max}=0.12$ (light lines) and 
$z_{\rm max}=0.15$ (dark lines) are considered.  
There is a negligible difference in the random measurement errors 
for these redshift limits. The correlation function is more accurately 
recovered for $z_{\rm max}=0.12$. 

The conclusions of this subsection are that an optimal measurement of the 
clustering of the 2PIGG catalogue will be obtained under the following 
conditions: 
(i) The pair counts in the NGP and SGP regions of the 
2dFGRS are combined. 
(ii) The value of $J_3$ is taken to be $1200$h$^{-3}$Mpc$^3$, though 
the results are fairly insensitive to the exact value. 
(iii) Cumulative bins in luminosity are used to define subsamples 
of groups from the 2PIGG catalogue. 
(iv)  $n_{\rm min}=2$ is used, which leads to the most reliable
measurements.
(v) The most accurate model of the selection function is obtained 
by a simple parametric fit to the observed group redshift distribution. 
(vi) A conservative redshift of $z_{\rm max}=0.12$ is used. 

\begin{figure*}
{\epsfxsize=16.truecm 
\epsfbox[30 20 570 560]{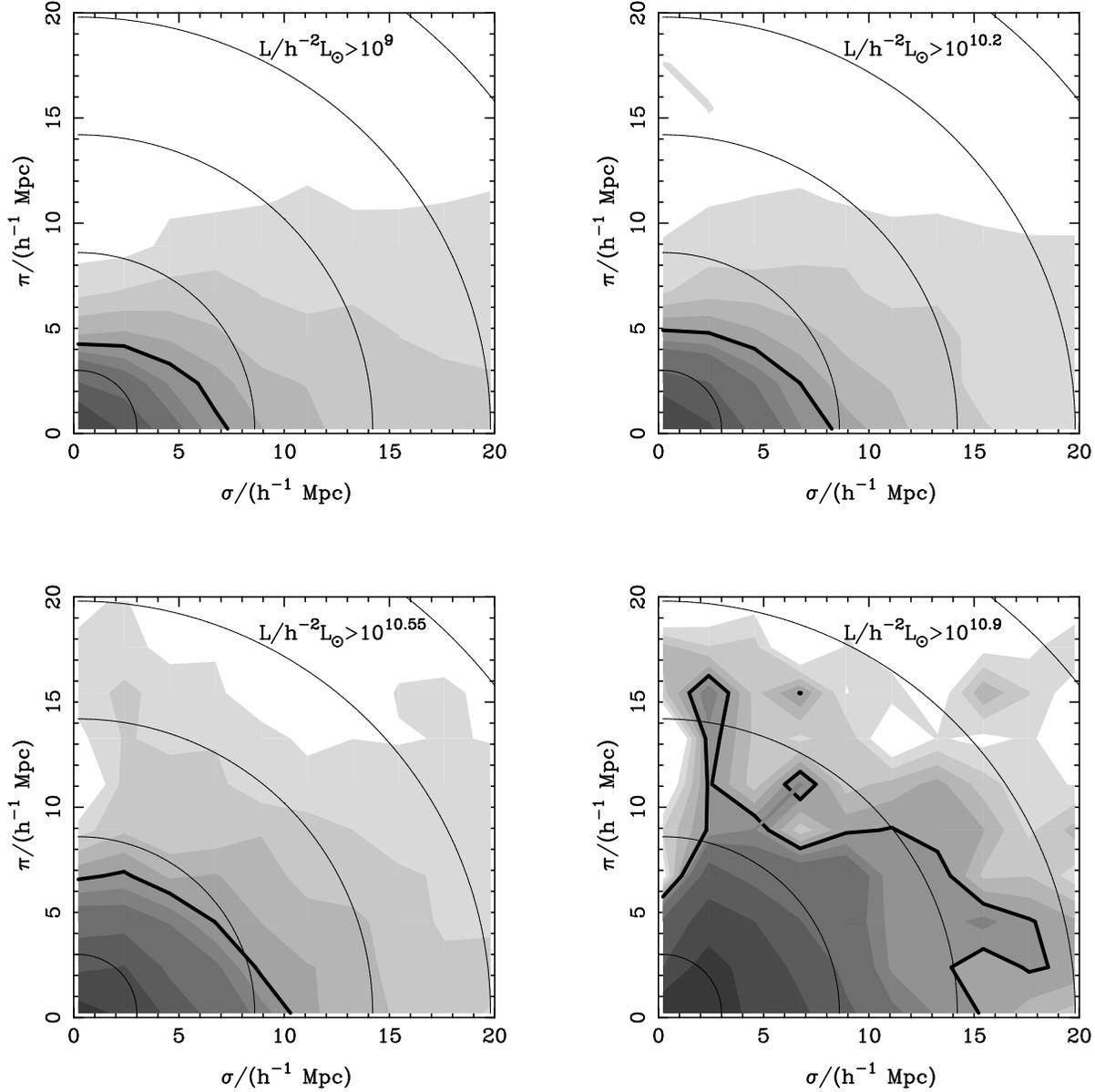}}
\caption{
The correlation function, $\xi(\sigma,\pi)$, for selected 2PIGG samples, 
as indicated by the label on each panel.
The grey scale indicates the amplitude of the correlation function, 
with darker shading indicating stronger clustering: 
$\xi(\sigma,\pi)=0.2,0.4,0.6,0.8,1,1.2,1.4,2,3$ and $5$ are shown.
The $\xi=1$ contour is also marked by a thick black line.
The thin black lines show the expected 
shape of the $\xi(\sigma,\pi)$ contours in the absence of redshift 
space distortions of the clustering pattern.
}
\label{fig:dist}
\end{figure*}

\subsection{The clustering of subsamples of groups}
\label{ssec:xissample}

We now investigate how well we can distinguish between the clustering 
signals displayed by different 
samples of groups defined by mass. We also test our 
clustering measurement algorithm to see how faithfully the clustering 
signal can reproduce an idealized measurement. This is an important 
consideration, as this determines the strength of the constraints we 
can place on theoretical models of the clustering of galactic systems. 

The errors on the measured correlation function are obtained from the 
{\it rms} scatter over the ensemble of Hubble Volume mock 2PIGG catalogues. 
We show these relative errors in Fig. \ref{fig:xierr}. 
The horizontal line marks an error of $100\%$. 
This plot indicates that we should be able to get a robust measurement of 
the correlation function out to pair separations of $s \sim 30h^{-1}$Mpc.
The small scale limit depends more upon the sample under 
consideration,  due to the changing space density of groups. 
For the full group sample, the correlation function can be measured down 
to separations below $s \sim 1h^{-1}$Mpc; for the most luminous sample, 
which has a much lower space density, the smallest pair 
separation that can be probed is $s \approx 3h^{-1}$Mpc.

To test the accuracy of our clustering measurements, 
we have extracted samples of groups from the {\tt GALFORM} mock 
catalogue defined by a series of lower limits in 
total group luminosity as discussed earlier. 
The clustering measured for these samples 
is compared with that of samples drawn from the simulation 
cube using two different 
criteria to construct equivalent samples as discussed earlier.  
The samples drawn from the simulation cube are idealized in the 
sense that they do not have a radial or angular selection function 
imposed upon them as is the case for the mock 2PIGG catalogue. 
Moreover, the groups occupy dark matter haloes identified by a 
friends-of-friends group finder.  The mass of a group in the 
simulation cube is known very accurately: it
is simply the sum of the mass of the dark matter 
particles connected by the group finder.

The correlation functions measured for mock group subsamples defined 
by a lower limit on total group luminosity are compared with the results 
for the equivalent samples drawn from the simulation 
cube in Fig. \ref{fig:ximockcube}. 
The top row of this figure shows the comparison when the equivalent samples 
are set up to reproduce the mass function of groups recovered from the mock 
group catalogue for each luminosity subsample. 
The bottom row shows the comparison for equivalent samples constructed to 
match the effective bias estimated from the mock subsamples.
The left hand column shows the results for groups with a minimum 
membership $n_{\rm min}=2$, and the right hand column for $n_{\rm min}=4$.
The clustering measurements from the mock subsamples are in impressively good 
agreement with the results obtained for the equivalent samples in the 
simulation cube, particularly for the case of $n_{\rm min}=2$.
The equivalent samples constructed to reproduce the effective bias of 
mock group subsamples are in somewhat better agreement with the measurements 
from the mocks for the brighter group samples.

The agreement in clustering amplitude between samples of groups 
taken from a mock and those extracted from an idealized simulation 
is significant; it means that we fully understand the impact of 
making a mock group catalogue on the measurement of the clustering of 
groups and can use the clustering of 2PIGG samples to constrain theoretical 
models of structure formation.

\begin{figure}
{\epsfxsize=8.truecm 
\epsfbox[20 140 589 715]{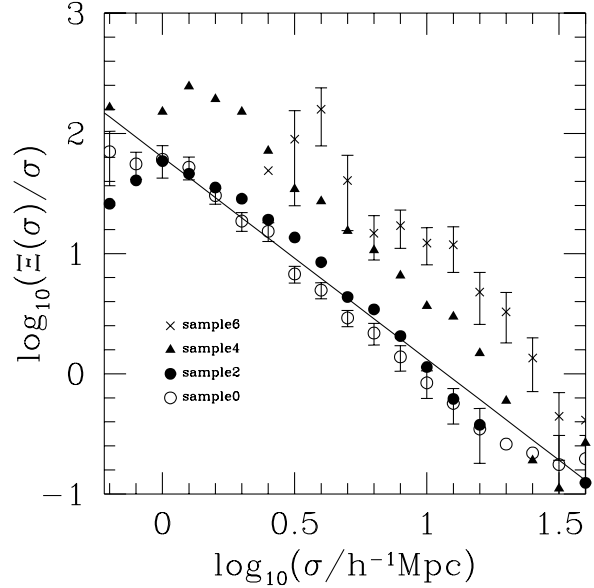}}
\caption{
The projected correlation function of selected 2PIGG samples.
The samples are indicated by the key; error bars are plotted on 
the measurements for selected samples for clarity.
The solid line shows a fit to the projected correlation function 
of 2dFGRS galaxies taken from Hawkins et al. (2003).
}
\label{fig:si4}
\end{figure}

\section{2PIGG results}

\begin{figure*}
\begin{picture}(450,250)
\put(0,0){\psfig{file=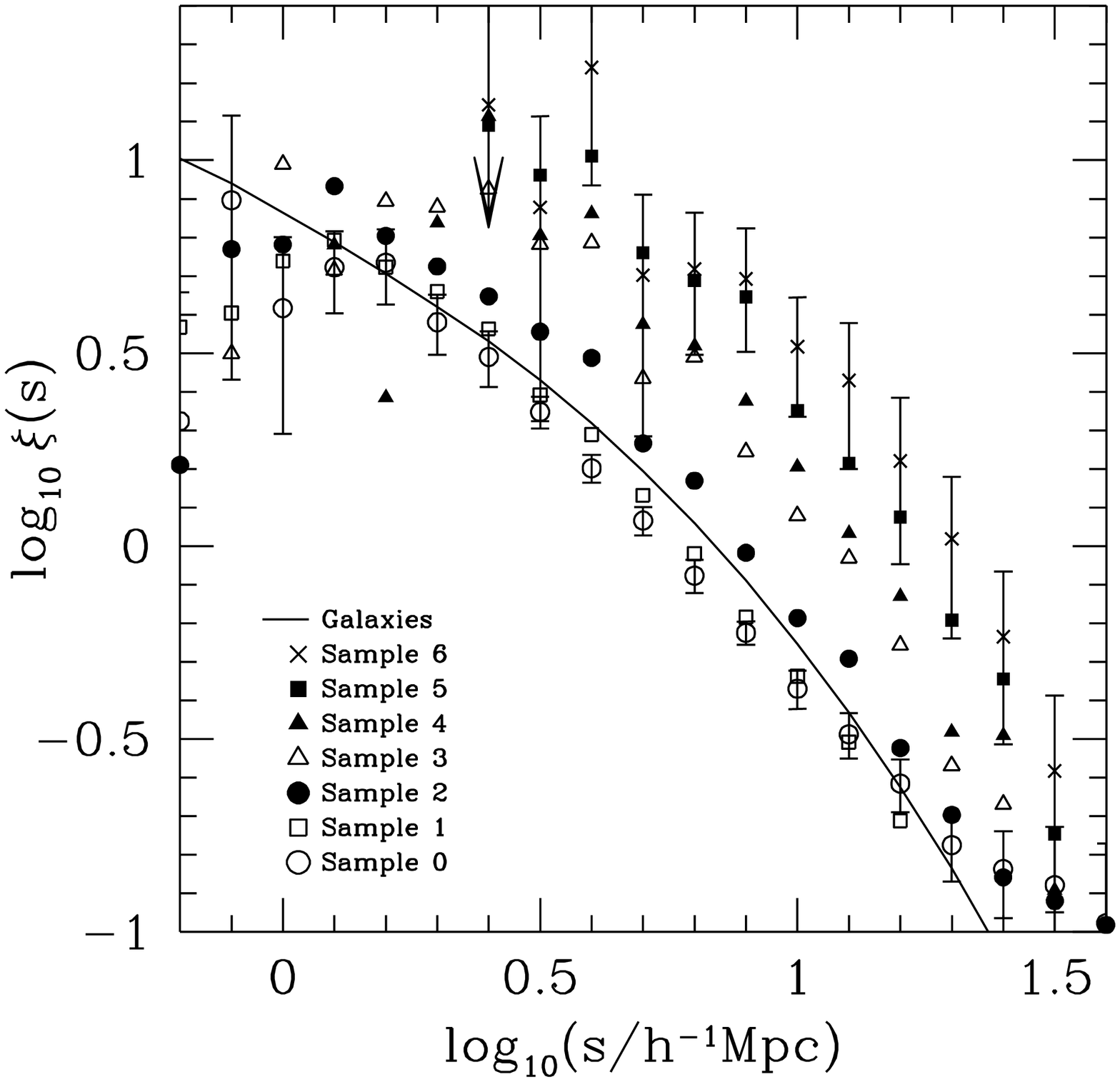,width=8cm,height=8cm}}
\put(225,0){\psfig{file=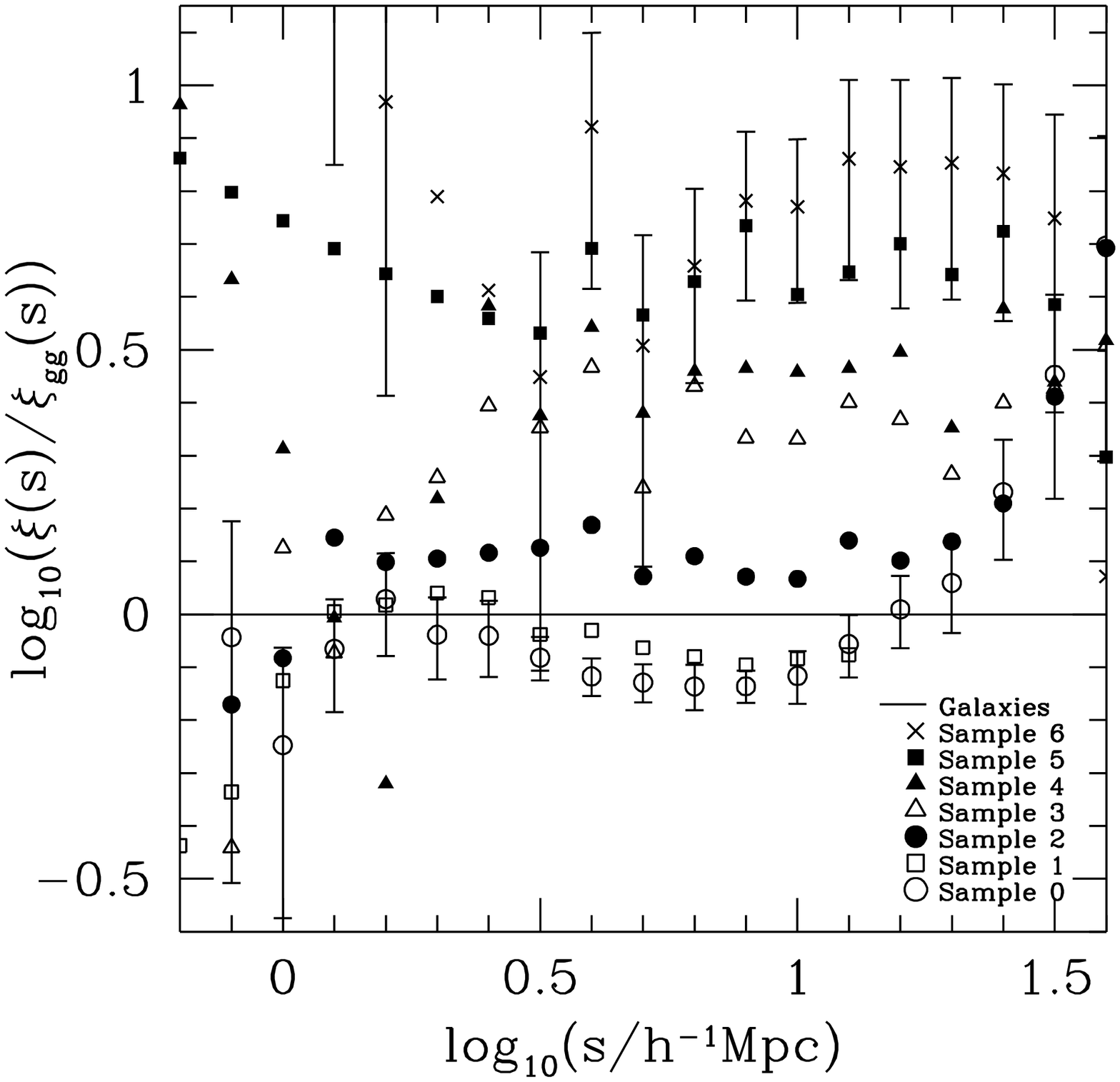,width=8cm,height=8cm}}
\end{picture}
\caption{
Left panel: 
The redshift space correlation function, $\xi(s)$, measured 
for samples of 2PIGGs. 
The samples are shown by the key. Error bars are plotted for selected 
samples for clarity.
The solid line shows $\xi(s)$ for 2dFGRS galaxies, measured by Hawkins 
et al. (2003).
Right panel:
The ratio of the group correlation function to the galaxy correlation 
function, plotted on a logarithmic scale. 
}
\label{fig:xi4}
\end{figure*}

In this section we measure the clustering amplitude of the real 2PIGGs 
for samples defined by total group luminosity. Some basic properties 
of the samples are given in Table 1. 
We apply the optimal clustering measurement 
algorithm set out at the end of Section \ref{ssec:optimal}. 
We combine the pair counts in these regions 
to estimate the correlation function. We have checked that the 
measurements for the individual regions agree within the errors.

An important test of the integrity of a group catalogue is 
the form of the correlation function measured in bins of pair 
separation parallel ($\pi$) and perpendicular ($\sigma$) to the 
line of sight, $\xi(\sigma,\pi)$.
Padilla \& Lambas (2003a,b) explored different techniques used in the 
construction of group and cluster catalogues. 
They found that the shape of $\xi(\sigma, \pi)$ is a powerful way to 
reveal spurious agglomerations of galaxies. 
A tell-tale sign of problems with the composition of galaxy groups 
is a significant enhancement of the correlation function, $\xi(\sigma, \pi)$, 
in the $\pi$ direction, as seen in the early analyses of Abell catalogue 
clusters (e.g. Bahcall \& Soneira 1983; Sutherland 1988). 
Such a distortion of the clustering pattern is seen for 2dFGRS galaxies, 
due to the large peculiar motions inside massive virialised 
structures (Peacock et al. 2001; Hawkins et al. 2003).
However, for the case of groups and clusters, such virialised 
structures are not predicted in viable models of structure 
formation (Kaiser 1987; Padilla et al. 2001; Padilla \& Baugh 2002). 
The correlation function, $\xi(\sigma, \pi)$, for samples of 2PIGGs is 
shown in Fig. \ref{fig:dist}. There is no enhancement of the clustering 
signal in the $\pi$ direction. The contours of constant 
clustering amplitude are, however, affected by peculiar motions, showing the 
flattening due to the infall motions expected in the gravitational 
instability paradigm (Padilla et al. 2001; Peacock et al. 2001). There is 
also a clear increase in clustering amplitude with sample luminosity. 

The projected correlation function, $\Xi(\sigma)/\sigma$, is the integral 
of $\xi(\sigma,\pi)$ over the $\pi$ direction (e.g. Norberg et al. 2001).
This statistic is unaffected by peculiar motions. 
In Fig. \ref{fig:si4}, we show the projected correlation functions for 
selected samples of 2PIGGs. The black line shows the best fit to the projected 
correlation function of galaxies in the 2dFGRS for comparison 
(Hawkins et al. 2003). 
We find that the lowest luminosity sample plotted is less strongly 
clustered than the galaxy distribution, a point to which we return later 
on in this section.
The brightest sample of groups displays a clustering 
amplitude that is an order of magnitude stronger than that measured for 
2dFGRS galaxies.  

The projected correlation function, due to the way it is calculated, can 
be unreliable if the correlation function is noisy on large scales. A more 
robust quantity in such cases is the redshift space correlation function, 
$\xi(s)$. This quantity is the average of $\xi(\sigma,\pi)$ 
within annuli centred on $s=\left(\sigma^2 + \pi^2\right)^{1/2}$. 
The redshift space correlation function measured for 2PIGG samples with 
membership $n_{\rm min} \ge 2$ is plotted in the left-hand panel of 
Fig. \ref{fig:xi4}. For reference, we also plot the redshift space 
correlation function of 2dFGRS galaxies measured by Hawkins et al. (2003). 
The correlation function of the group samples has a similar shape to that 
obtained for galaxies, on scales where a comparison is reliable. 
There is, however, a dramatic change in clustering amplitude with group  
luminosity, mirroring that seen for the projected correlation function 
in Fig. \ref{fig:si4}. These points are emphasized in the right 
hand panel of Fig. \ref{fig:xi4}, in which we plot the group 
correlation function divided by the galaxy correlation function. There is 
a relative bias between the spatial distribution of groups and galaxies. 
For the full 2PIGG sample, and also for the next faintest luminosity sample, 
this relative bias is actually an {\it anti-bias}; these groups are more 
weakly clustered than the galaxies. The brightest groups, on the other hand, 
are almost ten times more strongly clustered than the galaxy distribution.

A useful summary of the clustering measurements can be made by plotting 
the correlation length, $s_0$, as a function of the mean group separation, 
$d_c$, which is related to the space density of groups, $n$, 
by $d_{c}=1/n^{1/3}$. 
The space density of groups is estimated using the cumulative 
$b_{\rm J}-$band luminosity function of 2PIGGs out to 
$z=0.12$ (Eke et al. 2004b).  The space density of 
each subsample is simply the luminosity function evaluated at the 
appropriate lower limit in luminosity.
We estimate the correlation length (defined by $\xi(s_{0})=1$) by 
fitting a power law to the measured redshift space correlation function 
for pair separations in the range 
$0.4 <\log_{10}(s/$h$^{-1}$Mpc$)<1.3$. 
Our results are fairly insensitive to perturbations to this range. 

\begin{figure}
{\epsfxsize=8.truecm 
\epsfbox[ 20 160 570 700]{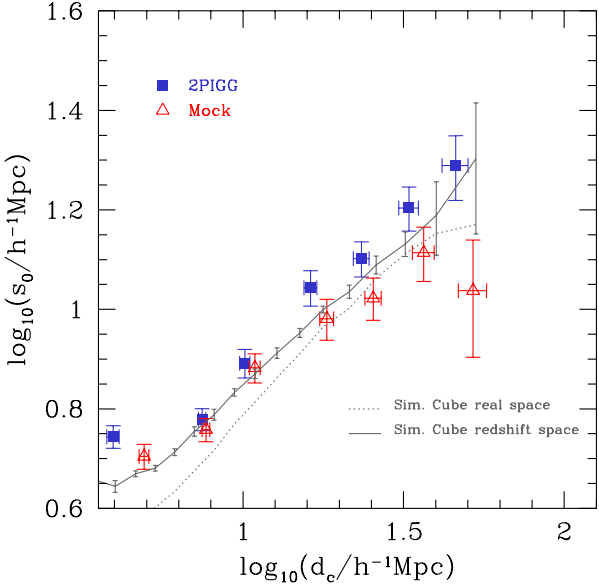}}
\caption{
The correlation length in redshift space, $s_0$, plotted as a function 
of the mean group separation, $d_c$.
The 2PIGG results are shown by filled squares for $n_{\rm min}=2$.
Also shown for comparison are measurements from the {\tt GALFORM} mock 
plotted with triangles. 
The solid line shows the $s_0-d_c$ relation obtained from the 
simulation cube for equivalent samples of groups (see text). The 
dotted line shows how these results change when the correlation 
length is measured in real space. 
}
\label{fig:s0dccomp}
\end{figure}

The clustering strength of 2PIGG samples versus mean group separation 
is plotted in Fig. \ref{fig:s0dccomp}. 
There is an increase in clustering strength with the mean separation 
of groups in each sample; the clustering strength increases slightly 
less rapidly than the change in group separation. 
There is very good agreement between the results for $n_{\rm min}=2$ and 
$n_{\rm min}=4$ for group separations for which a comparison is possible. 
Fig. \ref{fig:s0dccomp} also shows the $s_0$-$d_c$ relation 
in the {\tt GALFORM} mock 2PIGG catalogue with $n_{min}=2$, which is in 
very close agreement with the results from the real 2PIGG samples. 
The solid and dotted lines show the measurements for equivalent  
samples of groups drawn from the simulation cube. The solid line gives 
the $s_0$-$d_c$ relation in redshift space and the dotted line shows the 
real space results. The clustering amplitude is typically $10\%$-$20\%$ 
weaker in real space than in redshift space.

\begin{figure}
{\epsfxsize=8.truecm 
\epsfbox[ 10 140 580 715]{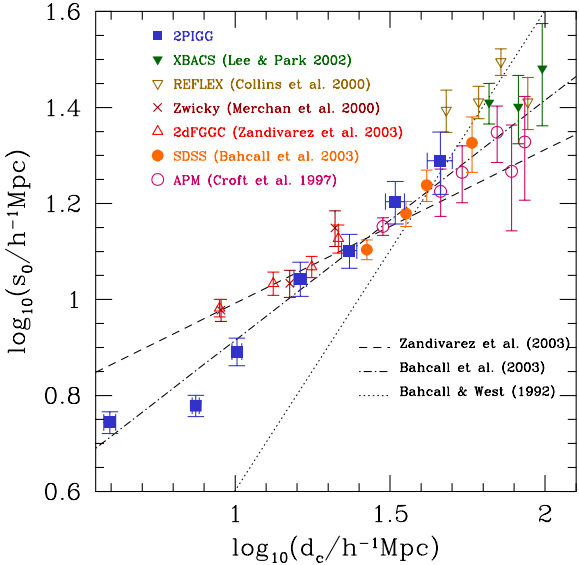}}
\caption{
A comparison of the $s_0-d_c$ relation for different samples.  
The squares show the results from the 2PIGGs for $n_{\rm min}=2$.
The other symbols show a selection of data taken from the literature, 
as indicated by the key.
The lines show fits to subsets of the data, 
with sources indicated by the key.
}
\label{fig:s0dcdata}
\end{figure}

\begin{figure}
{\epsfxsize=8.truecm 
\epsfbox[ 10 140 580 715]{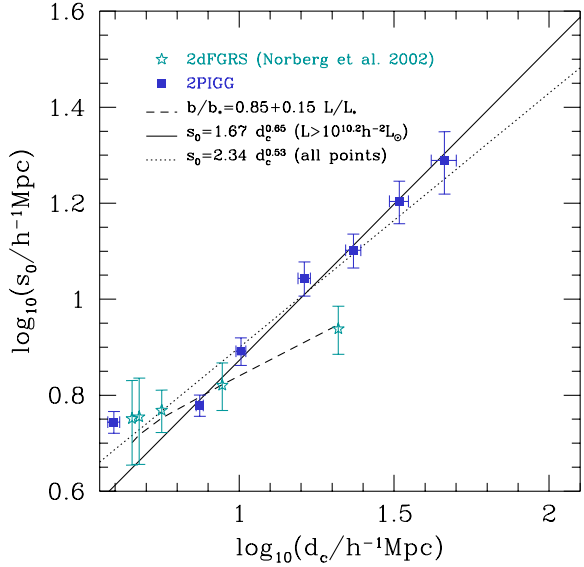}}
\caption{
A comparison of the $s_0-d_c$ relation for 2PIGG groups (squares) and 
for 2dFGRS galaxies (stars, taken from Norberg et al. 2002a).
The solid line is a fit to the 2PIGG clustering data, excluding the 
lowest luminosity sample; the dotted line is a fit to the 2PIGG 
results including the faintest sample. The dashed line 
shows the fit to the clustering strength-luminosity relation fitted to 
2dFGRS data by Norberg et al. (2001).  
}
\label{fig:s0dcgal}
\end{figure}

The 2PIGG results are compared with a selection of 
measurements taken from the literature in Fig. \ref{fig:s0dcdata}. 
This comparison between datasets should be treated with caution due 
to a number of differences in the way the various samples have been 
defined and analysed: 
namely, in decreasing order of seriousness:  
(i) the technique used to identify groups and clusters, 
(ii) the derivation of a value for the mean inter-group separation and 
(iii) the calculation of the correlation length.

Zandivarez et al. (2003) measured the clustering of groups in a 
catalogue constructed from the 100k release of 2dFGRS data by 
Merch\'{a}n \& Zandivarez (2002).
The two most abundant samples of groups analysed by 
Zandivarez et al. have a significantly higher clustering amplitude than 
we recover for 2PIGG groups of comparable abundance. 
Zandivarez et al. estimated $d_c$ from a dynamical 
estimate of the group mass which they translated into a spatial abundance 
using the measured mass function of groups (Mart\'{i}nez et al. 2002). 
Moreover, Zandivarez et al. only consider groups with $n_{\rm min}=4$; 
such groups are probably only associated with low mass systems through 
errors in the dynamical mass estimates. 
The abundance of groups in the Updated Zwicky Catalogue was estimated 
in a similar way by Merch\'{a}n et al. (2000).
Bahcall et al. (2003) have applied a photometric technique (Annis et al. 
2004) to a subset of the Sloan survey data to obtain a sample of 
clusters that overlaps with the most luminous 2PIGG groups and 
with APM Survey clusters (Croft et al. 1997; Dalton et al. 1997). 
The Sloan measurements are in reasonable agreement with the results 
from APM clusters and are marginally lower than the 2PIGG values at 
comparable abundances. However, both the SDSS and APM 
clusters are identified in projection.
The trend of clustering strength versus abundance found for 2PIGGs 
agrees quite well with the measurements from X-ray selected 
cluster samples (Abadi, Muriel \& Lambas 1998; 
Collins et al. 2000; Lee \& Park 2002).

The dotted line in Fig. \ref{fig:s0dcdata} shows the $s_{0}$-$d_{c}$ 
relation advocated by Bahcall \& West (1992), whereas the dot-dashed line 
is the relation proposed by Bahcall et al. (2003); the dashed line 
is the fit of Zandivarez et al. (2003) to their data. 
The analogous fit to the 2PIGG results is plotted in Fig. \ref{fig:s0dcgal} 
for clarity. If the faintest groups are ignored, the 2PIGG results 
are described by the relation $s_{0}=1.67d_{c}^{0.65}$. This is 
somewhat steeper than the fit of Bahcall et al. (2003).
 
We compare the clustering of 2PIGG groups with 2dFGRS galaxies in 
Fig. \ref{fig:s0dcgal}. The clustering amplitude of 2dFGRS galaxies 
is taken from Norberg et al. (2002a), who analysed volume limited 
samples defined by galaxy luminosity. The curve plotted over the 
galaxy clustering measurements shows the predictions of the simple 
relation between relative clustering strength and luminosity deduced  
by Norberg et al. (2001). 
The full sample of 2PIGGs is more weakly clustered than $L_*$ and 
brighter galaxies. On the other hand, the trend of clustering strength 
with pair separation is stronger for groups than it is for galaxies.

\section{Conclusions}
\label{sec:conc}

We have measured the clustering of groups in the 2PIGG sample 
constructed from the 2dFGRS by Eke et al. (2004a). The group catalogue, 
made up of galactic systems with a minimum of two members, contains 
$\sim$29,000 groups, out of which $\sim$16,000 are at $z<0.12$. 
The sample is sufficiently large and homogenous that a 
robust measurement of clustering is possible for subsamples of groups 
defined by their corrected total luminosity.
Our analysis has relied extensively on the use of various mock 
group catalogues constructed using high resolution N-body simulations 
populated with galaxies using the {\tt GALFORM} semi-analytic model 
(Cole et al. 2000; Benson et al. 2000, 2003).

In summary, the main conclusions we have reached are the following: 

\begin{enumerate}

\item
The main goals of this paper were to make a robust measurement 
of the clustering of galaxy groups with the aim of exploring 
the dependence of the clustering amplitude on a property 
related to group mass (for example the total group luminosity)  
and to compare the results with theoretical predictions.
We have tested our algorithm for estimating the correlation function 
by comparing results obtained from a mock 2PIGG catalogue with those 
derived from equivalent samples drawn from the N-body simulation cube. 
The success of this comparison is significant for two reasons. 
Firstly, in order to make a robust measurement of the clustering of groups 
from a flux limited galaxy survey, it is necessary to make a number of 
approximations and to choose certain parameter values (see Section 
\ref{ssec:optimal}). We have tuned the algorithm for measuring the 
clustering of 2PIGG groups by requiring a close match between the 
correlation functions estimated from the mocks and the original results from 
the simulation cube. Secondly, the ability to extract equivalent samples 
of groups from a simulation volume with the same clustering as samples 
taken from mock 2PIGG catalogues makes it possible to 
interpret the 2PIGG clustering results in the context of structure 
formation models.

\item
We find that clustering amplitude increases substantially with 
total group luminosity. 
The 2PIGG catalogue allows clustering measurements to be made for samples 
spanning a factor of $25$ in median total luminosity. The correlation function 
increases in amplitude 
by an order of magnitude over this luminosity interval. Another 
way of expressing this is that the redshift space correlation length changes 
by a factor $\approx 3.5$ from the faintest to the brightest sample. 
The most luminous groups we consider have a mean pair separation 
that is $\approx 5.5$ times greater than that of the full group sample.
There is little change in the shape of the correlation function with 
group luminosity. 

\item
The shape of the redshift space correlation function of groups is very 
similar to that measured for 2dFGRS galaxies on the scales for which 
a comparison is possible. The full 2PIGG sample has a {\it weaker} 
clustering amplitude than is measured for 2dFGRS galaxies; the 
correlation length of the 2PIGG sample with $n_{\rm min}=2$ is 
$s_0 = 5.5 \pm 0.1 h^{-1}$Mpc, whilst that of 2dFGRS galaxies is 
$6.82\pm0.28h^{-1}$Mpc (Hawkins et al. 2003). 
However, the clustering amplitude of brighter samples of groups is much 
greater than that of 2dFGRS galaxies.
 
\item
The correlation functions measured for 2PIGG samples are 
in very good agreement with the predictions of a semi-analytic 
model of galaxy formation, the {\tt GALFORM} model (Cole et al. 2000; 
Benson et al. 2003), in a cold dark matter universe with a 
cosmological constant ($\Lambda$CDM).
Previous work has compared the predictions of the $\Lambda$CDM model 
with measurements of the correlation length versus abundance relation 
for rich clusters (Governato et al. 1999; Colberg et al. 2000).
To extend the theoretical calculations down to group-sized systems, it 
is essential to extend the predictions of the spatial distribution 
of dark matter haloes in the $\Lambda$CDM cosmology with a model for 
galaxy formation. The galaxy formation model predicts how the 
dark matter haloes should be populated with galaxies. 
For this reason, the clustering of groups provides a more stringent test 
of theories of galaxies formation than the clustering of the richest clusters.

\item 
The trend of correlation length (measured in redshift space) against mean 
inter-group separation can be quantified as: $s_0 = 1.67 d_{c}^{0.65}$. 
The 2PIGG catalogue has made possible a robust measurement of 
the clustering of galactic systems, ranging from poor groups to 
rich clusters, from one sample for the first time. 
Our results are in general agreement with those obtained previously 
for rich groups and clusters.

\end{enumerate}

Galaxies and groups of galaxies trace the underlying distribution of 
dark matter haloes in different ways.
Galaxies trace the halo distribution in a complex way. 
The halo occupation distribution (HOD), which depends strongly on halo mass, 
 has been used extensively to constrain theoretical 
models (e.g. Benson et al. 2000; Peacock \& Smith 2000; Seljak 2000; 
Berlind et al. 2003).  
In principle, the group sample should trace the halo distribution in a  
simple fashion, with every halo above some mass  
spawning one galaxy group. 
However, in practice, 
the identification of groups in real surveys is complicated, making this 
ideal difficult to attain.
Because of this it is essential to devise a scheme to quantify the fidelity 
of a group catalogue and to uncover any biases in comparisons with 
theoretical models by applying the group finder to mock catalogues (Eke 
et al. 2004a).

The difference between the clustering amplitude measured for 
the full 2PIGG and the full 2dFGRS galaxy samples can be readily understood 
in terms of the HOD. Ideally, the groups have a one-to-one correspondence 
with the underlying dark matter haloes within some mass interval. 
The galaxies obviously sample haloes in the same mass range, 
but with a weight that increases with halo mass, since the most 
massive haloes host more than one galaxy. Thus, the more massive haloes, 
which are the more strongly clustered, make a larger contribution 
to the correlation function of 2dFGRS galaxies than they do 
in the case of the full 2PIGG sample.
The clustering of galaxy groups as a function of group luminosity therefore 
provides an important new test of theories of galaxy formation.

\section*{Acknowledgments}
This work was supported 
by a PPARC rolling grant at the University 
of Durham.  
NDP acknowledges receipt of a British
Council-Fundaci\'on Antorchas fellowship.
CMB was supported by a Royal Society University
Research Fellowship.

\end{document}